\documentclass[iop,revtex4]{emulateapj}

\slugcomment{DRAFT: \today}
\received{January 27, 2016}
\revised{}
\accepted{March 31, 2016}
\shorttitle{The MUSCLES Treasury Survey \mbox{Ly$\alpha$} and Extreme-UV Spectra}
\shortauthors{Youngblood et al.}

\usepackage{subfigure}
\usepackage{color}
\usepackage[hidelinks]{hyperref}

\hypersetup{ colorlinks, citecolor=black, filecolor=black, linkcolor=blue, urlcolor=blue }

\newcommand{\Lya}{\mbox{Ly$\alpha$}}

\newcommand{\HI}{H\,\textsc{i}}
\newcommand{\DI}{D\,\textsc{i}}

\newcommand{\MgII}{\ion{Mg}{2}}
\newcommand{\SiII}{\ion{Si}{2}}
\newcommand{\SiIII}{\ion{Si}{3}}
\newcommand{\SiIV}{\ion{Si}{4}}
\newcommand{\CII}{\ion{C}{2}}
\newcommand{\CIV}{\ion{C}{4}}
\newcommand{\HeII}{\ion{He}{2}}
\newcommand{\kms}{km s$^{-1}$}

\begin{document}

\title{The MUSCLES Treasury Survey II: Intrinsic Lyman Alpha and Extreme Ultraviolet Spectra of K and M Dwarfs with Exoplanets*
        }

\author{
               Allison Youngblood$^1$, Kevin France$^1$, R. O. Parke Loyd$^1$, Jeffrey L. Linsky$^2$, Seth Redfield$^3$, P. Christian Schneider$^4$, Brian E. Wood$^5$,  Alexander Brown$^6$, Cynthia Froning$^7$, Yamila Miguel$^8$, Sarah Rugheimer$^9$, Lucianne Walkowicz$^{10}$
                              	    }

\altaffiltext{*}{Based on observations made with the NASA/ESA Hubble Space Telescope, obtained from the data archive at the Space Telescope Science Institute. STScI is operated by the Association of Universities for Research in Astronomy, Inc. under NASA contract NAS 5-26555.}    
\altaffiltext{1}{Laboratory for Atmospheric and Space Physics, University of Colorado, 600 UCB, Boulder, CO 80309; allison.youngblood@colorado.edu}
\altaffiltext{2}{JILA, University of Colorado and NIST, 440 UCB, Boulder, CO 80309}
\altaffiltext{3}{Astronomy Department and Van Vleck Observatory, Wesleyan University, Middletown, CT 06459-0123}
\altaffiltext{4}{European Space Research and Technology Centre (ESA/ESTEC), Keplerlaan 1, 2201 AZ Noordwijk, The Netherlands}
\altaffiltext{5}{Naval Research Laboratory, Space Science Division, Washington, DC 20375}
\altaffiltext{6}{Center for Astrophysics and Space Astronomy, University of Colorado, 389 UCB, Boulder, CO 80309}
\altaffiltext{7}{Dept. of Astronomy C1400, University of Texas, Austin, TX, 78712}
\altaffiltext{8}{Laboratoire Lagrange, Universite de Nice-Sophia Antipolis, Observatoire de la Cote d'Azur, CNRS, Blvd de l'Observatoire, CS 34229, 06304 Nice cedex 4, France}
\altaffiltext{9}{Department of Earth and Environmental Sciences, Irvine Building, University of St. Andrews, St. Andrews KY16 9AL, UK}
\altaffiltext{10}{The Adler Planetarium, 1300 S Lakeshore Dr, Chicago IL 60605}

\begin{abstract}

The ultraviolet (UV) spectral energy distributions of low-mass (K- and M-type) stars play a critical role in the heating and chemistry of exoplanet atmospheres, but are not observationally well-constrained. Direct observations of the intrinsic flux of the \Lya~line (the dominant source of UV photons from low-mass stars) are challenging, as interstellar \ion{H}{1}~absorbs the entire line core for even the closest stars. To address the existing gap in empirical constraints on the UV flux of K and M dwarfs, the MUSCLES HST Treasury Survey has obtained UV observations of 11 nearby M and K dwarfs hosting exoplanets. This paper presents the \Lya~and extreme-UV spectral reconstructions for the MUSCLES targets. Most targets are optically inactive, but all exhibit significant UV activity. We use a Markov Chain Monte Carlo technique to correct the observed \Lya~profiles for interstellar absorption, and we employ empirical relations to compute the extreme-UV spectral energy distribution from the intrinsic \Lya~flux in $\sim$\,100 \AA~bins from 100\,--\,1170 \AA. The reconstructed \Lya~profiles have 300 \kms~broad cores, while \textgreater \,1\% of the total intrinsic \Lya~flux is measured in extended wings between 300 \kms~to 1200 \kms. The \Lya~surface flux positively correlates with the \MgII~surface flux and negatively correlates with the stellar rotation period. Stars with larger \Lya~surface flux also tend to have larger surface flux in ions formed at higher temperatures, but these correlations remain statistically insignificant in our sample of 11 stars. We also present \ion{H}{1}~column density measurements for 10 new sightlines through the local interstellar medium.

\end{abstract}

\section{Introduction} \label{sec:Introduction}

Ultraviolet (UV) photons control the upper atmospheric heating and chemistry of exoplanets.  For exoplanets around low-mass stars (spectral types K and M), characterization of the UV environment has only recently begun (e.g., \citealt{Walkowicz2008,France2013,Shkolnik2014}).  Most prior observations focused on flare stars \citep{Landsman1993,Ayres1995,Hawley2003,Osten2005}, and atmospheric models of cool stars have mostly focused on photospheric, Ca\,\textsc{ii} H \& K, and H$\alpha$~emission (\citealt{Husser2013}, but also see \citealt{Fuhrmeister2005}). UV observations of several optically inactive M dwarfs have shown that their chromospheres and transition regions are active and contribute significant flux to the UV spectral energy distributions \citep{France2013}. The very few previously available UV spectra of M dwarfs has limited our ability to accurately model the atmospheres and potential habitability of rocky planets around low-mass stars.  

M dwarfs are the most ubiquitous stellar type in the universe, outnumbering G stars 12 to 1 within 10 pc (The RECONS Team\footnote{www.recons.org}). Exoplanets in M dwarf habitable zones are easier to detect by both the transit and radial velocity methods, making targeted searches for habitable exoplanets around M dwarfs potentially more fruitful than for any other stellar type.  \cite{Dressing2015} estimated that there are 0.18\,--\,0.27 Earth-sized planets (1\,--\,1.5 $R_{\oplus}$) and 0.11\,--\,0.25 super-Earths (1.5\,--\,2 $R_{\oplus}$) per M dwarf habitable zone.

Some of the issues thought to threaten the habitability of an exoplanet orbiting a low-mass star are currently believed to be surmountable (e.g., \citealt{Tarter2007}). \cite{Joshi2003} showed that close-in, tidally locked exoplanets can maintain a habitable climate as long as significant surface pressure exists to support atmospheric heat transport to the nightside. The high-energy flux with which M dwarfs irradiate their exoplanets may be similar to the young Sun \citep{Lammer2011}. \cite{Joshi2003} noted that numerous starspots can cause the stellar luminosity to vary with rotation by approximately 40\%, but the same change is experienced by a planet with the eccentricity of Mars, and life may have been possible on Mars \citep{Noffke2015}. Unresolved issues include the effects of particle fluxes (enhanced during flares and coronal mass ejections) on the magnetospheres and atmospheres (e.g., \citealt{Lammer2007}; \citealt{Segura2010}; \citealt{Cohen2014}), UV-driven water loss \citep{Luger2015}, and UV-driven photochemistry in terrestrial and gaseous exoplanet atmospheres \citep{Grenfell2014,Miguel2014,Rugheimer2015}.

Observations of potential biosignatures (e.g., O$_2$, O$_3$, and CH$_4$) and habitability indicators  (e.g., H$_2$O and CO$_2$) in these planets' atmospheres will likely be possible in the next two decades (e.g., 2015 AURA High-Definition Space Telescope report\footnote{http://www.hdstvision.org/release-text}). However, UV-driven photochemistry can mimic certain biosignatures, complicating the analysis \citep{Harman2015,Tian2014,Domagal-Goldman2014}. \Lya~and other bright chromospheric and transition region resonance lines in the far-UV (912\,--\,1700 \AA) photodissociate CO$_2$ and H$_2$O, resulting in a buildup of abiotic O$_2$. O$_2$ dissociation by 1200\,--\,2400 \AA~photons leads to a three-body reaction involving O and O$_2$ that accumulates abiotic O$_3$ in the exoplanet's atmosphere. Because these reactions depend on the relative balance of far-UV to near-UV (1700\,--\,3200 \AA) flux, characterizing the full stellar UV spectrum is critical for atmospheric modeling to interpret the origin of biosignatures.

Low-mass stars have weak near-UV and far-UV continua punctuated by emission lines from the chromosphere and transition region, including \ion{H}{1}~Lyman alpha (\Lya; $\lambda$1216 \AA). \Lya~comprises $\sim$37\%\,--\,75\% of the total 1150\,--\,3100 \AA~flux from most M dwarfs \citep{France2013}, but is significantly attenuated by optically thick \ion{H}{1}~absorption from the intervening interstellar medium (ISM), even for the closest stars ($d$\,\textless \,22 pc in this study). The intrinsic \Lya~emission profile must be reconstructed from the observed profile (e.g., \citealt{Wood2005,France2013,Linsky2013}) to properly characterize the far-UV spectrum. 

The \Lya~flux is also one proxy used to estimate the extreme-UV (100\,--\,912 \AA) flux, which heats exoplanetary upper atmospheres, driving exoplanet mass-loss \citep{Lammer2003,Lammer2007,Lammer2009}.  The extreme-UV is heavily attenuated for all stars, except the Sun, by neutral hydrogen in the ISM from 912 \AA~to approximately 400 \AA, where ground-state neutral hydrogen's photoionization cross-section becomes an order of magnitude smaller. No astronomical instrument currently exists to observe the 100\,--\,400 \AA~spectral window.

An unknown intrinsic profile shape and complicated ISM structure along the line of sight are challenges when reconstructing the intrinsic \Lya~profiles of low-mass stars. The Sun's intrinsic \Lya~profile is hundreds of \kms~broad with narrow self-absorption at the peak (e.g., \citealt{Curdt2008,Woods1995,Fontenla1988}), and its \MgII~profile has a similar shape \citep{Purcell1963}. For K dwarfs, \cite{Wood2005} assumed a \Lya~ profile similar to the ISM absorption-corrected \MgII~line, which exhibits self-reversal. M-dwarf temperature-pressure profiles may be different enough from the Sun's that self-reversal may not be present. Sunspots, which have similar effective temperatures to M dwarfs, do not show \Lya~self-reversal \citep{Fontenla1988,Curdt2001,Tian2009}, and M-dwarf \MgII~profiles do not show self-reversal \citep{Wood2005}. Also, \Lya's linewidth is expected to be much narrower than the Sun's due to the positive correlation between stellar luminosity and linewidths of optically-thick chromospheric lines \citep{Wilson1957,Ayres1979,Cassatella2001}.

The $Hubble$ $Space$ $Telescope$ MUSCLES\footnote{Measurements of the Ultraviolet Spectral Characteristics of Low-mass Exoplanetary Systems} Treasury Survey has directly measured the 1150\,--\,3200 \AA~spectra of 11 nearby ($d$\,\textless \,22 pc) K and M dwarf exoplanet hosts in order to provide an empirical basis for stellar-irradiance estimates. ``Paper I" \citep{France2016} provides an overview of the science goals and observing strategy of the MUSCLES survey as well as a summary of initial results. In this ``Paper II" of a series from the MUSCLES team, we present the reconstructions of the intrinsic \Lya~stellar emission lines and the extreme-UV irradiance. ``Paper III" \citep{Loyd} discusses the data reduction steps followed to merge spectra from the $HST$ observations and other sources into panchromatic spectral energy distributions (SEDs) and examines the photodissociation of several common molecules when directly exposed to these SEDs. In this work, Sections~\ref{sec:ObservationsReductions} and~\ref{sec:Methodology} describe the \Lya~observations, reductions, and the methodology we use to reconstruct the intrinsic \Lya~profiles for the stars in the survey. Section~\ref{sec:Results} presents the \Lya~and extreme-UV reconstructions that have been incorporated into panchromatic spectra of our targets (available as High Level Science Products on MAST\footnote{https://archive.stsci.edu/prepds/muscles/}). We compare \Lya~with other chromospheric emission lines and discuss the profile shape in Section~\ref{sec:Discussion} and summarize our results in Section~\ref{sec:Summary}.

\section{\Lya~Observations and Reductions} \label{sec:ObservationsReductions}

Observations were obtained between October 2014 and August 2015 with the Space Telescope Imaging Spectrograph (STIS) and Cosmic Origins Spectrograph (COS) on the $Hubble$ $Space$ $Telescope$ ($HST$) (see \citealt{France2016} for an overview of the observing strategy). To observe \Lya, we used the STIS G140M grating (R\,$\sim$\,11,400) with the 52\arcsec~$\times$~0.1\arcsec~slit for the seven M dwarfs and the E140M grating (R\,$\sim$\,48,500) with the 0.2\arcsec~$\times$~0.06\arcsec~slit for the four K dwarfs. The small slit sizes were chosen to minimize geocoronal \Lya~airglow contamination. The absolute velocity accuracy near \Lya~for G140M is 2.5\,--\,6.2 \kms~and for E140M is 0.7\,--\,1.6 \kms. 

Although STIS's narrow slit minimizes airglow photons entering the spectrograph, geocoronal \Lya~can still dominate the stellar \Lya~signal. The \ion{H}{1}~column densities of the local ISM are large enough ($N$(\ion{H}{1})\,$\gg$\,14) to completely attenuate the \Lya~line cores for the stars in

\begin{figure*}[t]
   \begin{center}
   
     \subfigure{
          \includegraphics[width=\textwidth]{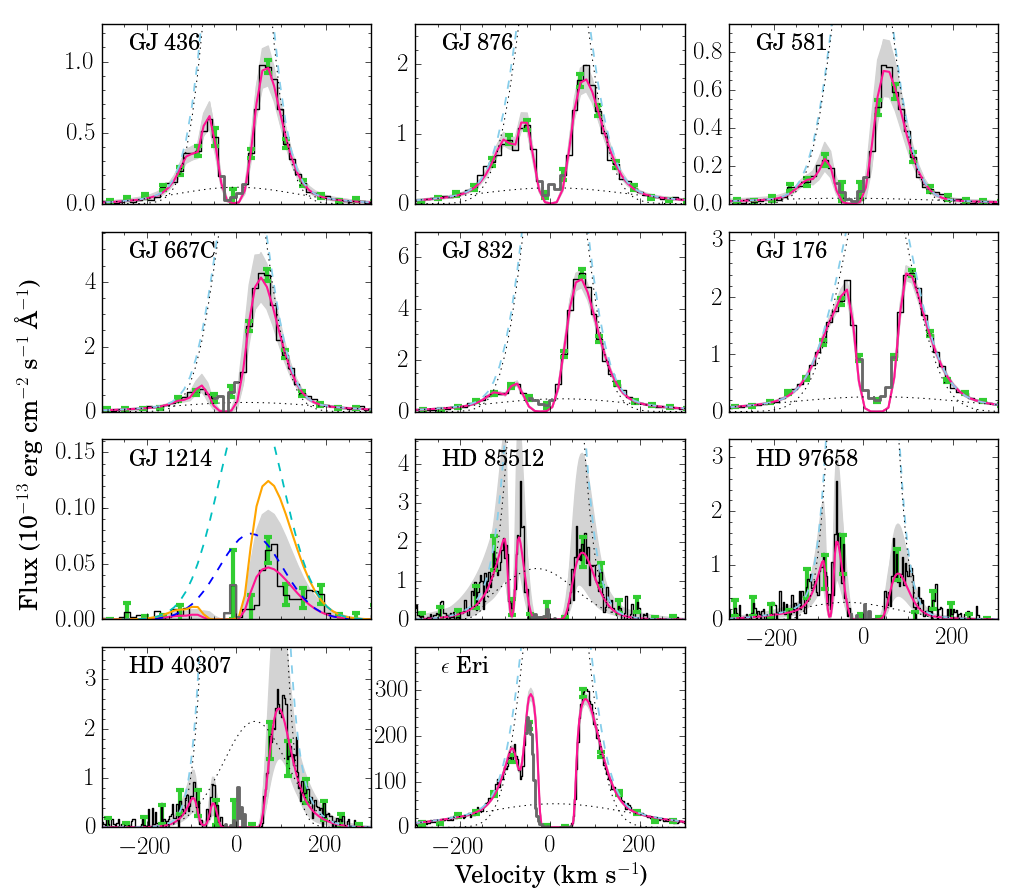}
          }
   \end{center}
    \caption{%
        MCMC solutions for the 7 M dwarfs and 4 K dwarfs. The black histograms with green error bars show STIS G140M or E140M data, the pink lines are the best-fit attenuated profiles determined by MCMC, the gray-shaded regions show the uncertainty of the fits, and the dashed blue lines are the reconstructed intrinsic \Lya~profiles. The dotted lines show the narrow and broad components of the reconstructed profiles. The sections of data masked from the fits are shown with the dark gray histogram. In GJ 1214's panel, the solid pink line, gray-shaded region, and dashed, dark blue line represent the original MCMC fit, and the solid orange and dashed, light blue line represent the scaled solution (see Section~\ref{sec:GJ1214}).}
    \label{fig:all_MCMC}

\end{figure*}

our sample (log$_{10}$ $F$(\Lya)\,$\approx$ \,-14 -- -10 erg cm$^{-2}$ s$^{-1}$). To further minimize geocoronal \Lya~contamination of the observed \Lya~profiles, we scheduled each target's observations during the time of year when Earth's heliocentric velocity would allow the geocoronal \Lya~airglow to coincide with the attenuated stellar line core. Airglow was subtracted using the detector regions outside the spectral trace. Residual airglow is resolved in the E140M observations, allowing for a second airglow subtraction from the one-dimensional spectra. The G140M observations do not have sufficient resolution for the second iteration of airglow subtraction, so we mask the attenuated line core from the fits (see Section~\ref{sec:Methodology}), because the line cores should be consistent with zero flux for all of our targets. To avoid biasing the \Lya~line profile fits to the low-signal wings, we ensured the error bars were greater than or equal to the rms scatter in the wings over the the velocity range -300 to -400 \kms.

Because COS's aperture is large (2.5\arcsec), the COS G130M observations are too contaminated by airglow to access the \Lya~line core ($\pm$\,300 \kms), but they do detect the broad wings of \Lya~(see Section~\ref{sec:lineprofile}). We refer the reader to \cite{Loyd} for details on the reduction of the STIS and COS observations and the construction of the panchromatic SEDs.

\section{Methodology} \label{sec:Methodology}

\subsection{The Model} \label{sec:TheModel}
To reconstruct the intrinsic \Lya~profiles from the observed profiles (presented fully in Section~\ref{sec:Results}), we fit a 9 parameter model describing the intrinsic stellar \Lya~profile (6 parameters) and the interstellar \ion{H}{1}~and \ion{D}{1}~absorption (3 parameters) via a Markov Chain Monte Carlo (MCMC) technique (see Section~\ref{sec:MCMC}). Figure~\ref{fig:all_MCMC} shows the observed \Lya~profiles (black histograms with green error bars). In $\epsilon$~Eri's panel, the \ion{H}{1}~and \ion{D}{1}~absorption are seen around 0 \kms~and -75 \kms, respectively. The fits shown in Figure~\ref{fig:all_MCMC} will be discussed in detail in Section~\ref{sec:Results}.

\begin{figure}[t]
   \begin{center}
   
     \subfigure{
          \includegraphics[width=0.5\textwidth]{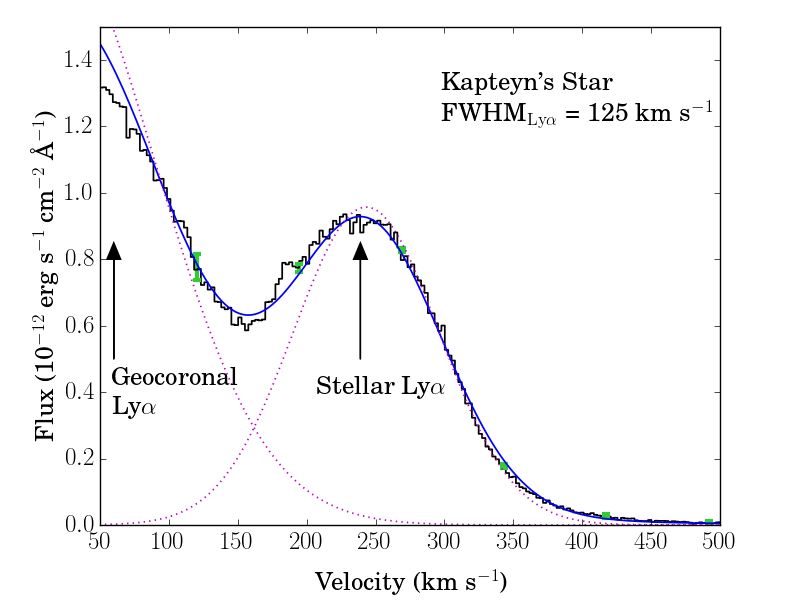}
          }
   \end{center}
    \caption{
        The COS G130M spectrum \citep{Guinan2016} of Kapteyn's star (spectral type sdM1.0) is shown with the black histogram and green error bars. The blue line is the best simultaneous fit to the wing of the geocoronal \Lya~and stellar \Lya~emission, with magenta dotted lines showing the individual components of the fit before convolution with the wavelength-dependent COS line-spread function.  The FWHM of Kapteyn's Star's \Lya~profile is 125 \kms.}
    \label{fig:Kapteyn}

\end{figure}  

We assume for the intrinsic \Lya~profile a dominant narrow Gaussian and weaker broad Gaussian, each described by an amplitude $A$ (erg cm$^{-2}$ s$^{-1}$ \AA$^{-1}$), FWHM (\kms), and heliocentric velocity centroid $V$ (\kms). The \Lya~line has not been modeled from first principles for any M dwarf, and semi-empirical models provide only a rough fit to the data \citep{Fontenla2016}. As discussed in Section~\ref{sec:Introduction}, no self-reversal of the line core is expected. We searched the $HST$ archive for a \Lya~observation of a high radial-velocity M dwarf whose stellar emission is Doppler shifted away from the geocoronal and ISM line centers. We analyzed the $HST$--COS spectrum of Kapteyn's Star (\citealt{Guinan2016}; GJ 191; spectral type sdM1.0). Figure~\ref{fig:Kapteyn} shows its Gaussian-shaped \Lya~profile with no self-reversal, redshifted ($V$ = +245 \kms) well away from the geocoronal emission and ISM absorption. This supports our choice of a double-Gaussian as the intrinsic \Lya~profile shape for the M dwarfs.

We assume for the K dwarfs the same profile shape as for the M dwarfs. \cite{Wood2005} used the absorption-corrected \MgII~profiles, which display self-reversal, as a \Lya~template. In the absence of a \MgII~template, \cite{Bourrier2013} used two Voigt profiles of equal width and damping constants separated in velocity as the intrinsic \Lya~profile. The velocity separation of the two emission components can mimic self-reversal of the line core. We discuss the effect of excluding self-reversal on the K dwarf \Lya~reconstructions in Section~\ref{sec:noselfreversal}.

To model the ISM \ion{H}{1}~and \ion{D}{1}~absorption, we assume a single Voigt profile parameterized by \ion{H}{1}~column density $N$(\ion{H}{1}) (cm$^{-2}$), Doppler $b$ value (\kms), velocity centroid $V_{\rm \HI}$ (\kms), and D/H ratio fixed at 1.5\,$\times$\,10$^{-5}$ in accord with the mean value found by \cite{Wood2004} for the local ISM.  We assume that thermal line broadening determines the Doppler width parameter ($b_{\rm \DI}$ = $b_{\rm \HI}$/$\sqrt{2}$), \ion{D}{1}~has the same velocity centroid as \ion{H}{1}, and the \ion{D}{1}~column density scales with the \ion{H}{1}~column density by the D/H ratio. Local ISM conditions are fairly well known (e.g., \citealt{Redfield2008}), and we expect Doppler $b$ values around 10 \kms~and log$_{10}$ \ion{H}{1}~column densities around 18 for nearby stars \citep{Wood2005}.

\begin{figure*}[t]
     \begin{center}
        \subfigure{%

            \includegraphics[scale=0.45]{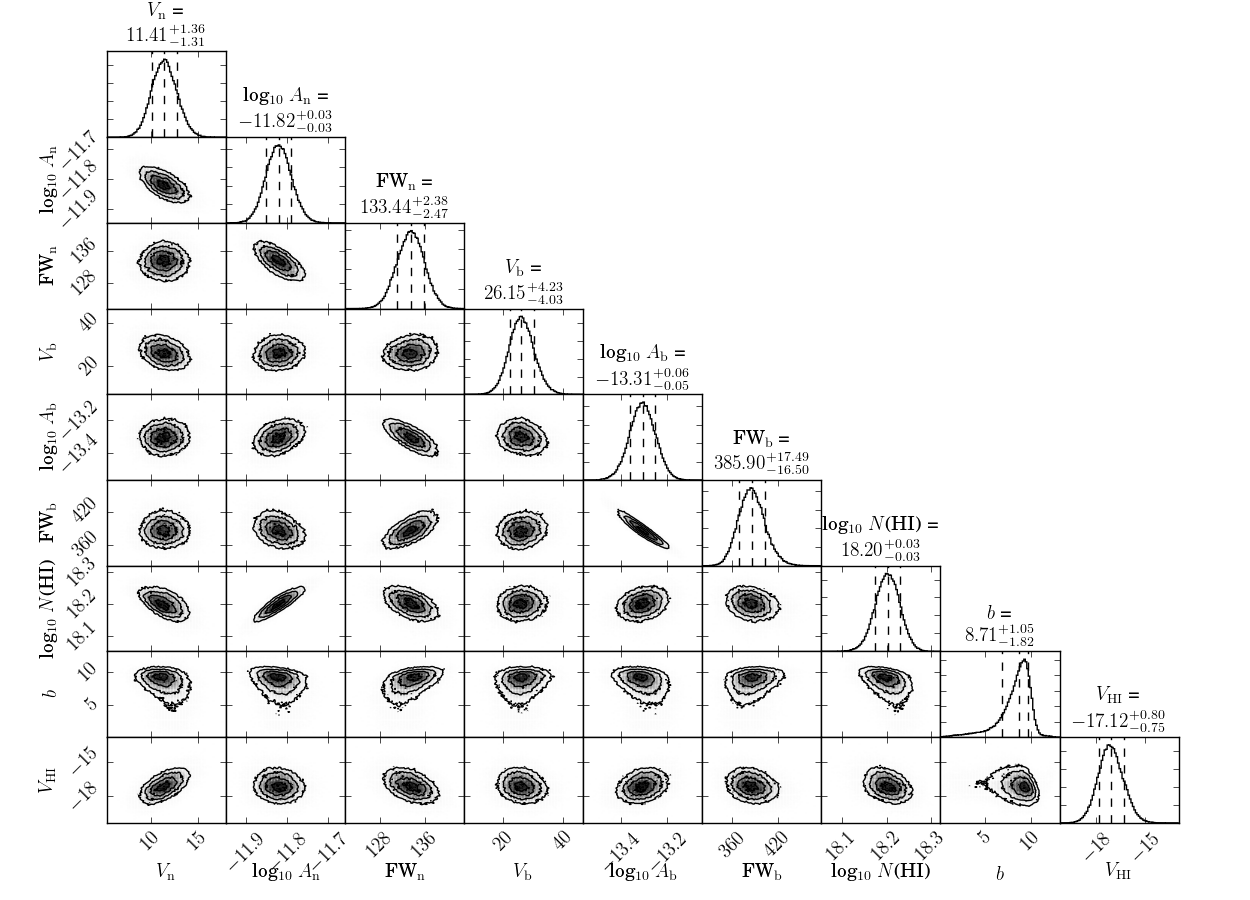}
        \label{fig:GJ832_cornerplot}
        }
        \subfigure{
            \includegraphics[scale=0.45]{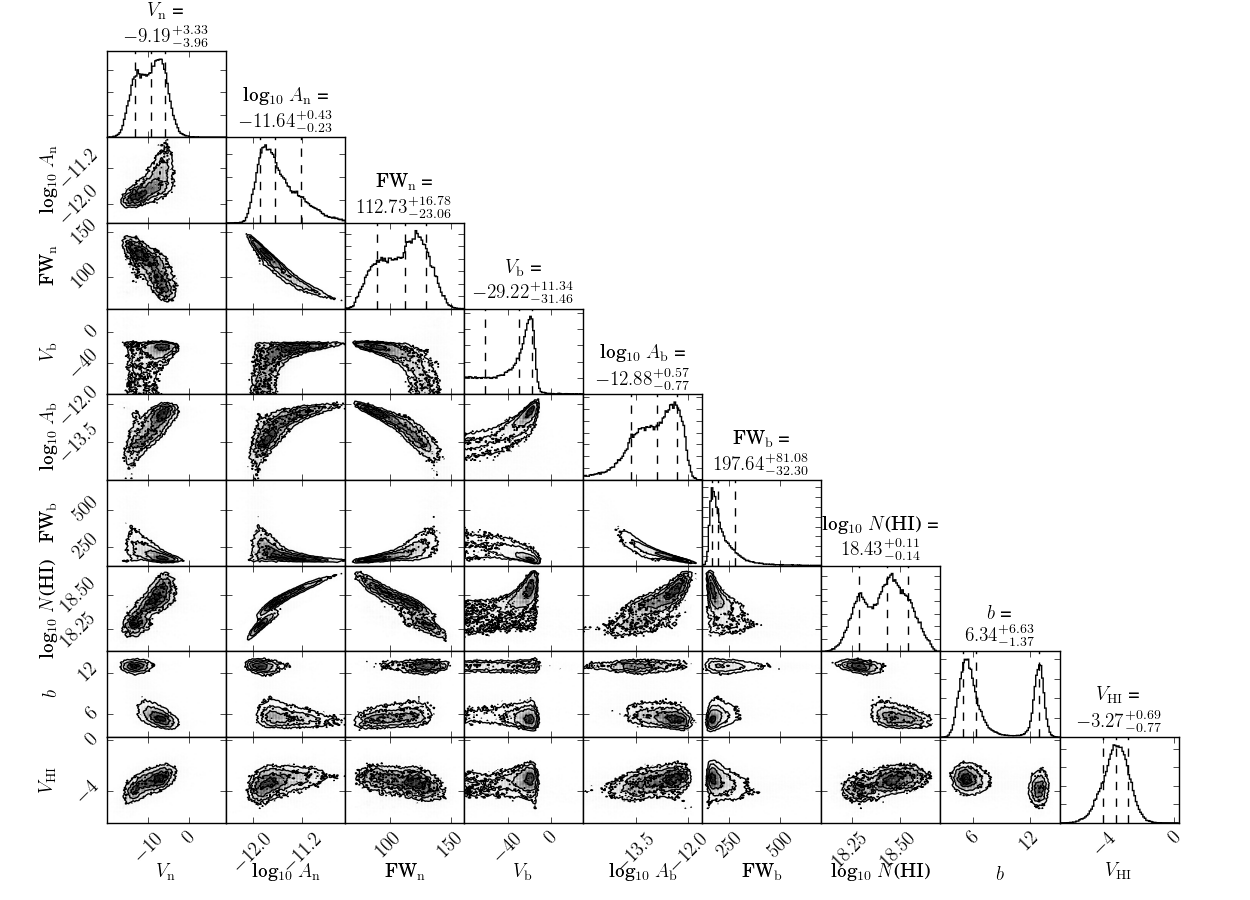}
        \label{fig:HD85512_cornerplot}
}

    \end{center}
    \caption{%
       One- and two-dimensional projections of the sampled posterior probability distributions, referred to as marginalized and joint distributions, respectively, of the 9 parameters for GJ 832 (top) and HD 85512 (bottom). Contours in the joint distributions are shown at 0.5-, 1-, 1.5-, and 2-$\sigma$, and the histograms' dashed vertical lines show the 16th, 50th, and 84th percentiles of the samples in each marginalized distribution. Similar figures for the other nine targets are available in the online version.
     } (The complete figure set (eleven images) is available in the online journal.)
   \label{fig:cornerplots}
\end{figure*}

\begin{figure*}[t]
   \begin{center}
   
     \subfigure{
          \includegraphics[width=\textwidth]{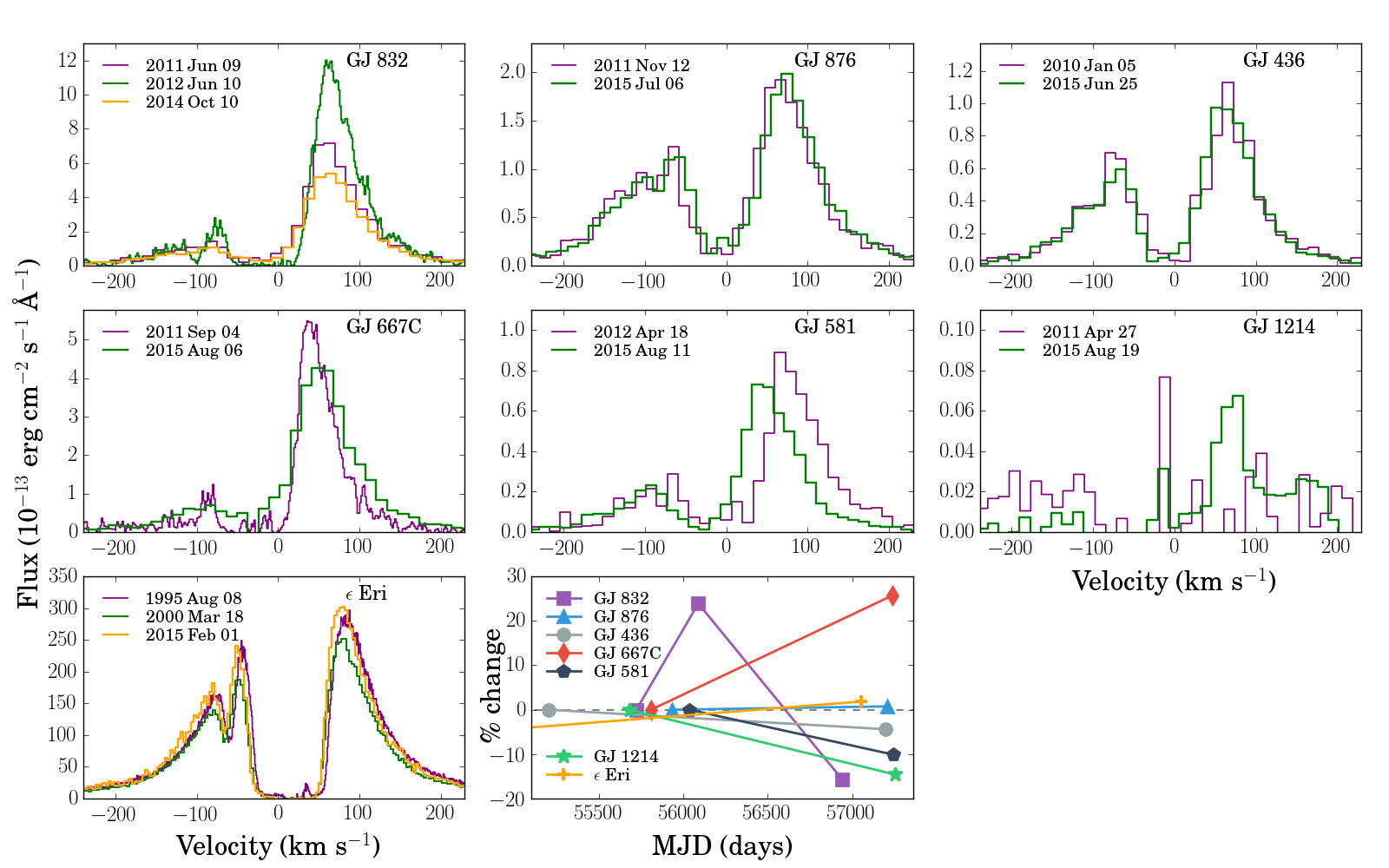}
          }
   \end{center}
    \caption{%
        The first 7 panels show multi-epoch \Lya~profiles observed with $HST$ STIS (the 1995 observation of $\epsilon$~Eri was with GHRS). The last panel shows the percent changes in the observed \Lya~fluxes. GJ 1214's 0\% value is based on an upper limit (Section~\ref{sec:GJ1214}). For $\epsilon$~Eri, the year 2000 observed flux decreased by 14\% relative to its 1995 value. The velocity shift between GJ 581's two epochs is likely due to an error in the wavelength calibration.
        }
    \label{fig:GJ832_comparison}

\end{figure*}

\subsection{MCMC} \label{sec:MCMC}

We employ the affine-invariant Markov Chain Monte Carlo (MCMC) sampler of \cite{Goodman2010} via the Python module \texttt{emcee} \citep{Foreman-Mackey2013} to efficiently sample the posterior distribution of our 9 parameters (see Table~\ref{table:res} for a list of the parameters). Several of the parameters (e.g., amplitude and column density) are expected to be correlated, and the affine-invariance of \texttt{emcee} ensures that the sampling is unaffected by parameter correlations. We assume a Gaussian likelihood and uniform priors for all parameters except the Doppler $b$ parameter, which has a logarithmic prior probability. We sample the posterior probability distribution 500,000 times for each target with a typical burn-in period of 300 steps and use one- and two-dimensional projections of the posterior probability distributions of the 9 parameters (referred to as marginalized and joint distributions, respectively) to analyze the fits. Figure~\ref{fig:GJ832_cornerplot} exemplifies the sampled posterior distributions of our high-S/N data and Figure~\ref{fig:HD85512_cornerplot} represents that of the low-S/N data. Similar figures for the other nine targets are available in the online journal.

Most of the marginalized distributions of the individual parameters are singly peaked, and the joint distributions exhibit the degeneracies between parameters (diagonally-aligned contours in Figure~\ref{fig:cornerplots}). The strongest degeneracies are between the narrow component's amplitude ($A_{\rm narrow}$) and log$_{10}$ $N$(\ion{H}{1}) and between the broad component's amplitude ($A_{\rm broad}$) and FWHM$_{\rm broad}$. Constant intrinsic flux can be maintained by simultaneously increasing (decreasing) $A_{\rm narrow}$ and increasing (decreasing) log$_{10}$ $N$(\ion{H}{1}), but the degeneracy between $A_{\rm broad}$ and log$_{10}$ $N$(\ion{H}{1}) is much weaker. There are also several parameter degeneracies that exhibit the trade-off between having a large or small integrated-flux contrast between the narrow and broad components. Other degeneracies involve the effects of having the velocity centroid of the ISM absorption ($V_{\rm \HI}$) offset from or coincident with the emission peak.

For each of the 9 parameters, the best-fit values are taken as the 50th percentile (the median) of the marginalized distributions, and 1-$\sigma$~error bars as the 16th and 84th percentiles (shown as dashed vertical lines in Figure~\ref{fig:cornerplots}). The best-fit reconstructed \Lya~fluxes are determined from the best-fit amplitude, FWHM, and velocity centroid parameters, and the 1-$\sigma$~error bars of the reconstructed \Lya~flux are taken by varying these parameters individually between their 1-$\sigma$~error bars and keeping all others fixed at their best-fit value.  The resulting minimum and maximum \Lya~fluxes become the 1-$\sigma$ error bars (Table~\ref{table:LyAfluxes}).

\subsection{Systematic effects} \label{sec:ModelAssumptions}
Here we evaluate potential systematic uncertainties not accounted for by the 1-$\sigma$ uncertainties determined by the MCMC method. We explore the effects of excluding \Lya~self-reversal, assuming a fixed D/H ratio, assuming a single ISM absorption component, and instrumental spectral resolution on the reconstructed \Lya~fluxes and the MCMC-derived 1-$\sigma$ uncertainties. 

The true \Lya~flux is not known for any of our target stars. Therefore, we carried out a number of simulations to confirm that our \Lya~reconstruction method works and investigated systematic uncertainties potentially not accounted for by the 1-$\sigma$~uncertainties determined by the MCMC method. Specifically, we simulated G140M and E140M data assuming different S/N ratios, one to three ISM components, and different D/H ratios. Fitting our model (one ISM component, fixed D/H ratio, no self-reversal) to these data, we find that the reconstructed \Lya~flux is always reasonably close to the true \Lya~flux and that for all targets except $\epsilon$~Eri, the uncertainties printed in Table~\ref{table:LyAfluxes} are close to 1-$\sigma$.

We have determined that fits to data with higher instrumental spectral resolution results in 1-$\sigma$ uncertainties that do not fully account for systematic uncertainties from the exclusion of self-reversal and additional ISM absorption components (see Sections~\ref{sec:noselfreversal} and \ref{sec:multicomp} for more details). We find that each can introduce up to a 30\% uncertainty. To show the worst case scenario where these uncertainties are present (where appropriate) and at their maximum, we add them here in quadrature with the MCMC 1-$\sigma$ uncertainties from Table~\ref{table:LyAfluxes}. We add a 30\% uncertainty term for excluding self-reversal for all four K dwarfs and another 30\% term for HD 85512 and HD 97658 for excluding additional ISM components. Both sightlines are believed to have 2 or 3 ISM components \citep{Redfield2008}. $\epsilon$~Eri's error bars increase dramatically ($\pm$\,3\% to $\pm$\,30\%), and the other three K dwarfs' increase modestly (roughly $^{+110}_{-45}$\% to $^{+115}_{-60}$\%). Note that we present the MCMC-determined 1-$\sigma$ uncertainties in the reconstructed \Lya~fluxes in Table~\ref{table:LyAfluxes}; these additional systematic terms are not included.

For the G140M observations, we have found some evidence for a $\sim$\,15\% systematic underestimate of the true intrinsic flux. However, this underestimate is usually less than the 1-$\sigma$ error bars on the reconstructed fluxes, so we do not attempt to correct this effect (see Section~\ref{sec:resolution} for more details). 

\subsubsection{Excluding self-reversal} \label{sec:noselfreversal}
Our model assumes that the K-dwarf intrinsic profiles do not show self-reversal of the \Lya~line core. While it has not been definitively shown that K stars' \Lya~profiles are self-reversed, previous authors \citep{Wood2005,Bourrier2013} fit self-reversed profiles to observed K dwarf profiles, so we wish to estimate the effect of excluding self-reversal. We apply our model and methodology to the STIS E140H spectra (R$\sim$114,000) presented in \cite{Wood2005} for the early K stars $\epsilon$~Eri, HR 8, and HR 1925. The MCMC reconstructed \Lya~fluxes for HR 8 and HR 1925 are $\sim$20\% smaller than \cite{Wood2005}'s values.  For both stars we find \ion{H}{1}~column densities 15--20\% smaller, thus our reconstructed fluxes are also smaller (see discussion of degeneracies in Section~\ref{sec:MCMC}).  For $\epsilon$~Eri, we find a \ion{H}{1}~column density 20\% larger than \cite{Wood2005} and \cite{Dring1997}, thus we find a 27\% larger reconstructed \Lya~flux. Excluding self-reversal appears to introduce up to a 30\% difference in the reconstructed \Lya~fluxes.

\subsubsection{Variable D/H Ratio} \label{sec:d2h}
For the results presented in the next section, we keep the D/H ratio fixed at 1.5$\times$10$^{-5}$ in accord with average value for the local ISM \citep{Wood2004}, but here we explore the effect of allowing the D/H ratio to be a free parameter with a uniform prior probability. When we allow the D/H ratio to vary (excluding GJ 1214 because its low S/N provides no constraint on the D/H ratio), we find an average D/H ratio in the MUSCLES sample of (1.4\,$\pm$\,0.6)\,$\times$\,10$^{-5}$. While this result agrees with \cite{Wood2004}, the D/H ratio in these fits was typically not well constrained and had large uncertainties. Except for $\epsilon$~Eri, the intrinsic \Lya~fluxes found while floating the D/H ratio all have solutions that are well within the 1-$\sigma$~error bars of the intrinsic fluxes found from the fixed D/H ratio fits. $\epsilon$~Eri's D/H ratio was found to be 2.2$\times$10$^{-5}$ and its reconstructed \Lya~flux was 13\% smaller.

Simulating G140M and E140M observations with different ``true" D/H ratios (1.3\,--\,1.7\,$\times$\,10$^{-5}$) introduces a dispersion in the reconstructed \Lya~fluxes that is smaller than the dispersion seen from simulating the observations repeatedly. Because the reconstructed \Lya~flux does not appear to depend sensitively on the assumed D/H ratio, and there is no evidence that the D/H ratio of the local ISM varies significantly from 1.5\,$\times$\,10$^{-5}$, we assume the effect of a fixed D/H ratio on the reconstructed \Lya~fluxes is negligible.

\subsubsection{Multiple absorption components} \label{sec:multicomp}
Since some of our sight lines are known to cross multiple interstellar clouds with different line of sight velocities \citep{Redfield2008}, we explore the effect of assuming a single ISM absorption component for all targets by adding a second absorption component to our fits for GJ 832, GJ 176, and HD 85512 ($d$ = 4.9 pc, 9.4 pc, and 11.2 pc).  

\cite{Redfield2008}'s dynamical model of the local ISM\footnote{http://lism.wesleyan.edu/LISMdynamics.html} has only one known interstellar cloud along GJ 832's line of sight (the Local Interstellar Cloud (LIC) with predicted radial velocity -11.3 \kms~and predicted log$_{10}$ $N$(\ion{H}{1}) = 16.6).  Two clouds are predicted along GJ 176's (the LIC with radial velocity 23.7 \kms~and log$_{10}$ $N$(\ion{H}{1}) = 18.2, and the Hyades cloud with radial velocity 12.9 \kms) sightline. Two clouds are also predicted along HD 85512's sightline (the G cloud with predicted radial velocity -0.1 \kms, and the Cet cloud with predicted radial velocity 14.3 \kms). Predicted column densities for the LIC come from \cite{Redfield2000}'s LIC model\footnote{http://lism.wesleyan.edu/ColoradoLIC.html}. There are no \ion{H}{1}~column density predictions for any other local cloud.

For all three targets, we find the second ISM component to have log$_{10}$ $N$(\ion{H}{1}) \textless~16, insignificant compared to the column densities of the first components (17.5 -- 18.5). For GJ 832, we find the two ISM components have log$_{10}$ $N$(\ion{H}{1}) = 18.2 and 14.4 and $V_{\rm \HI}$ = -17.1 \kms~and -18.6 \kms. The reconstructed \Lya~flux changed by \textless \,2\%. For GJ 176, the two ISM components have log$_{10}$ $N$(\ion{H}{1}) = 17.5 and 14.6 and $V_{\rm \HI}$ = 29.2 \kms~and 4.4 \kms. The reconstructed \Lya~flux changed by \textless \,2\%. For HD 85512, the two components have log$_{10}$ $N$(\ion{H}{1}) = 18.3 and 15.7 and $V_{\rm \HI}$ = -2.8 \kms~and -5.4 \kms. Although the quality of the fit did not change, the reconstructed \Lya~flux decreased by 40\%. The single-component and two-component fits are consistent within 1-$\sigma$~error bars for all three stars.

We also simulated and fitted G140M and E140M observations of sightlines with 1\,--\,3 ISM absorption components and found that for the G140M observations, the MCMC-derived error bars are approximately 1-$\sigma$ regardless of the number of ISM components along the sightline. For the E140M simulated observations with 2 and 3 ISM components, the MCMC-derived error bars may underestimate the true uncertainty. Depending on the ISM parameters and their velocity centroids compared to the intrinsic emission velocity centroids, we find over- and under-estimates of 10\,--\,30\% of the true intrinsic flux. 

We conclude that our low-resolution G140M and low signal-to-noise (S/N) E140M \Lya~profiles do not provide enough constraints on a second absorption component to justify its inclusion. We have shown in this section that for the G140M observations, there appears to be no systematic effect from excluding additional ISM components in the fits. For the E140M observations, the uncertainty introduced by ignoring a second or third ISM component could be as large as 30\%. The two E140M-observed stars in the MUSCLES sample with multiple ISM components along the line of sight are HD 85512 and HD 97658.

\subsubsection{Resolution effects} \label{sec:resolution}

We also explore the effects of resolution on our fits, because additional absorption features (multiple ISM components and astrospheric/heliospheric absorption) that are resolved in the E140M observations (R\,$\sim$\,45,800) will not be resolved in the G140M observations (R\,$\sim$\,11,400). We expect that the fitted parameters' uncertainties for high S/N G140M spectra should be larger than for high S/N E140M spectra, because resolving the \ion{D}{1}~absorption provides a tight constraint on the \ion{H}{1}~column density and velocity centroid, thereby reducing degeneracies in the model. 

We choose three stars ($\epsilon$~Eri, AU Mic, and 70 Oph A) with high S/N E140M observations from the $HST$ archive and smooth the data to simulate a G140M observation. We smooth the \Lya~profile to the approximate resolution of the G140M grating (R\,$\sim$\,11,400), and increase the error bar values to approximate the typical S/N per pixel of a G140M observation ($\sim$\,20). To quantify the absolute change in error bar size and test the stability of the best-fit parameters over many low-resolution ``observations" with random noise, we add random noise (drawn from Gaussian distributions with varying standard deviation proportional to the error bars) and use our MCMC technique to fit the simulated observations 100 times.

For all three stars, we retrieve reconstructed \Lya~fluxes $\sim$18\% smaller than what we find from the original E140M spectra. The simulated G140M spectra have reconstructed profiles with larger FWHMs and smaller amplitudes for both the broad and narrow Gaussian components. $A_{\rm narrow}$ and $A_{\rm broad}$ on average decreased by 22\% and 37\%, respectively, and FWHM$_{\rm narrow}$ and FWHM$_{\rm broad}$ increased by 13\% and 20\%, respectively. Because the total intrinsic flux is proportional to $A_{\rm narrow}$\,$\times$\,FWHM$_{\rm narrow}$ + $A_{\rm broad}$\,$\times$\,FWHM$_{\rm broad}$, these changes propagate to a 18\% decrease in flux. \ion{H}{1}~column densities decreased by 7\% on average, while Doppler $b$ values increased by only a few percent. The marginalized error bars for the individual parameters and the reconstructed \Lya~fluxes increase to levels consistent with the G140M observations ($\sim$\,20\%), although the stability of the reconstructed flux solution over the 100 trials is better ($\sigma$\,$\sim$\,10\%).

We list the specific findings for the three stars in Table~\ref{table:res}. For reference, $\epsilon$~Eri is a K dwarf with a single ISM velocity component and a detected astrosphere (the equivalent of the heliosphere around another star; \citealt{Wood2005}). AU Mic is an M dwarf with a single ISM velocity component and no detected astrospheric or heliospheric absorption, and 70 Oph A is a K dwarf with three known ISM velocity components and detected astrospheric and heliospheric absorption \citep{Wood2005}. For 70 Oph A, we use only one velocity component for the ISM absorption and make no attempt to treat the astrospheric and heliospheric absorption. At lower resolution, astrospheric absorption blends into the blueward wall of the \ion{H}{1}~absorption, and thus the fitted \ion{H}{1}~velocity centroid parameter shifts a few \kms~blueward and the Doppler $b$ increases slightly.

We conclude there likely is a 10\,--\,20\% systematic underestimate of the reconstructed fluxes for the G140M observations compared to the E140M observations, but note from our simulated observations that this underestimate is smaller than the 1-$\sigma$ error bars about 68\% of the time. We therefore do not correct our G140M results for this systematic error.

We also call attention to the result that the derived $N$(\ion{H}{1}) and $b$ values do not change appreciably when the spectral resolution is degraded to that corresponding to the G140M observations. This indicates that the $N$(\ion{H}{1}) and $b$ values derived from analysis of the G140M spectra are accurate. We also note that from our simulated observations with multiple ISM components, we find that the fitted $N$(\ion{H}{1}) parameter approximately equals the sum of the true ISM components' column densities.

\begin{deluxetable*}{lcc|cc|cc}
\tablecolumns{7}
\tablewidth{0pt}
\tablecaption{ Effect of resolution on the MCMC solution parameters$^{*}$ \label{table:res}} 
\tablehead{\colhead{} & 
                  \multicolumn{2}{c}{$\epsilon$~Eri} & 
                  \multicolumn{2}{c}{AU Mic} & 
                  \multicolumn{2}{c}{70 Oph A} \\ 
                  \colhead{ } & 
                  \colhead{High res.} & 
                  \colhead{Low res.} \vline & 
                  \colhead{High res.} & 
                  \colhead{Low res.} \vline & 
                  \colhead{High res.} & 
                  \colhead{Low res.} 
                  }
\startdata

$F$(\Lya) $\times$~10$^{-11}$\dotfill & 6.1$^{+0.2}_{-0.2}$ & 5.2$^{+1.3}_{-1.0}$ & 1.07$^{+0.04}_{-0.04}$ & 0.9$^{+0.2}_{-0.2}$ & 3.1$^{+0.1}_{-0.1}$ & 2.7$^{+0.7}_{-0.5}$ \\
(erg cm$^{-2}$ s$^{-1}$) &  &  &  &  & &  \\[4pt]
$^{\dagger}$$V_{\rm narrow}$\dotfill & 13.6$^{+0.4}_{-0.4}$ & 12.7$^{+1.5}_{-1.5}$ & -9.7$^{+0.7}_{-0.7}$ & -10.3$^{+1.7}_{-1.5}$ & -13.0$^{+0.4}_{-0.4}$ & -12.0$^{+2.2}_{-2.0}$ \\
(\kms) &  &  &  &  & &  \\[4pt]
$^{\dagger}$$A_{\rm narrow}$ $\times$~10$^{-11}$\dotfill & 9.4$^{+0.4}_{-0.4}$ & 7.1$^{+1.4}_{-1.1}$ & 1.2$^{+0.06}_{-0.06}$ & 0.9$^{+0.1}_{-0.1}$ & 5.1$^{+0.2}_{-0.2}$ & 4.2$^{+0.8}_{-0.6}$ \\
(erg cm$^{-2}$ s$^{-1}$ \AA$^{-1}$) &  &  &  &  &  &  \\[4pt]
$^{\dagger}$FWHM$_{\rm narrow}$\dotfill & 129.3$^{+1.6}_{-1.6}$ & 143.6$^{+5.8}_{-5.8}$ & 148.6$^{+2.7}_{-2.6}$ & 180.0$^{+11.2}_{-6.4}$ & 126.8$^{+0.8}_{-0.9}$ & 134.0$^{+4.3}_{-4.5}$ \\
(\kms) &  &  &  &  &  &  \\[4pt]
$^{\dagger}$$V_{\rm broad}$\dotfill & 11.7$^{+1.8}_{-1.8}$ & 11.6$^{+9.5}_{-9.3}$ & -8.5$^{+1.6}_{-1.5}$ & -4.8$^{+5.1}_{-4.8}$ & -4.9$^{+1.5}_{-1.6}$ & -5.4$^{+6.9}_{-6.7}$ \\
(\kms) &  &  &  &  &  &  \\[4pt]
$^{\dagger}$$A_{\rm broad}$ $\times$~10$^{-11}$\dotfill  & 0.52$^{+0.28}_{-0.26}$ & 0.34$^{+0.07}_{-0.06}$ & 0.19$^{+0.01}_{-0.01}$ & 0.10$^{+0.01}_{-0.03}$ & 0.21$^{+0.01}_{-0.01}$ & 0.15$^{+0.03}_{-0.02}$ \\
(erg cm$^{-2}$ s$^{-1}$ \AA$^{-1}$) &  &  &  &  &  &  \\[4pt]
$^{\dagger}$FWHM$_{\rm broad}$\dotfill & 400.1$^{+9.0}_{-8.9}$ & 477.9$^{+40.6}_{-35.0}$ & 387.8$^{+9.5}_{-8.7}$ & 493.2$^{+25.7}_{-23.3}$ & 378.4$^{+7.3}_{-7.2}$ & 431.0$^{+29.7}_{-26.2}$ \\
(\kms) &  &  &  &  &  &  \\[4pt]
$^{\ddagger}$log$_{10}$ $N$(\ion{H}{1})\dotfill & 17.93$^{+0.02}_{-0.02}$ & 17.88$^{+0.10}_{-0.11}$ & 18.35$^{+0.1}_{-0.1}$ & 18.32$^{+0.03}_{-0.04}$ & 18.28$^{+0.01}_{-0.01}$ & 18.23$^{+0.06}_{-0.06}$ \\
(cm$^{-2}$) &  &  &  &  &  &  \\[4pt]
$^{\ddagger}$$b$\dotfill & 12.1$^{+0.02}_{-0.02}$ & 13.8$^{+0.4}_{-0.7}$ & 14.0$^{+0.2}_{-0.2}$ & 14.1$^{+0.8}_{-0.8}$ & 15.5$^{+0.1}_{-0.1}$ & 15.7$^{+0.7}_{-1.1}$ \\
(\kms) &  &  &  &  &  &  \\[4pt]
$^{\ddagger}$$V_{\rm \HI}$\dotfill  & 16.1$^{+0.4}_{-0.4}$ & 10.0$^{+0.7}_{-0.5}$ & -24.0$^{+0.2}_{-0.3}$ & -26.8$^{+0.9}_{-1.0}$ & -31.2$^{+0.1}_{-0.1}$ & -32.6$^{+1.0}_{-1.0}$ \\
(\kms) &  &  &  &  &  &  \\

\enddata
\tablenotetext{*}{Error bars on the individual parameters are marginalized over all other parameters.}
\tablenotetext{$\dagger$}{\Lya~emission parameters. }
\tablenotetext{$\ddagger$}{ISM absorption parameters. }
\end{deluxetable*}

\subsection{Comparison of fitting techniques} \label{sec:ComparisonMethods}

To verify the robustness of our MCMC optimizations, we also fit our model using MPFIT \citep{Markwardt2009} via the Python package \texttt{pyspeckit} \citep{Ginsburg2011}. All reconstructed intrinsic \Lya~fluxes agree within the MCMC results' error bars, although the individual parameters can vary more widely.

We also fit the M0 V star AU Mic with our MCMC technique to ascertain the agreement of our fitting technique and model with that of \cite{Wood2005} for a star not expected to show self-reversal of the \Lya~line core. Our result shows excellent agreement with \cite{Wood2005}'s reconstructed \Lya~flux and the ISM absorption parameters within a few percent. We find a total flux $F$(\Lya) = (1.07$\pm$0.04) $\times$~10$^{-11}$ erg cm$^{-2}$ s$^{-1}$, log$_{10}$ $N$(\ion{H}{1}) = 18.35$\pm$0.01, $b$ = 14.0$\pm$0.2 \kms, and $V_{\rm \HI}$ = -24.0$^{+0.2}_{-0.3}$ \kms. \cite{Wood2005} found $F$(\Lya) =  1.03$\times$10$^{-11}$ erg cm$^{-2}$ s$^{-1}$, log$_{10}$ $N$(\ion{H}{1}) = 18.356$\pm$0.002, $b$ = 13.74$\pm$0.08 \kms, and $V_{\rm \HI}$ = -20.83$\pm$0.09 \kms.

\subsection{Time variability} \label{sec:TimeVariability}

The observed \Lya~flux of our targets changes over time. We do not see any evidence for \Lya~flaring in our STIS time-tag light curves, but multiple epochs of observations show some large changes. Figure~\ref{fig:GJ832_comparison} shows a comparison of the observed \Lya~profiles for the seven targets with multiple observations and quantifies the changes in observed flux over time. GJ 832, GJ 667C, and $\epsilon$~Eri exhibit significant changes in observed flux, and the other four stars are consistent with no change.

Between 2011 and 2012, the observed (not reconstructed) \Lya~flux of GJ 832 increased by $\sim$\,25\%, and by 2014, the observed flux decreased to $\sim$\,15\% below the 2011 level. GJ 667C's observed profile appears broader in 2015 than it did in 2011, even after convolving the 2011 observed profile to the approximate resolution of the 2015 observation. The observed flux increased 25\%, but we note the 2011 observation has low S/N. For $\epsilon$~Eri, the \Lya~profiles (including the astrospheric absorption) are similar in the 1995 and 2015 observations, but they show a decrease of 14\% in 2000.

GJ 876, GJ 436, and GJ 581 (assuming the apparent blueshift from 2012 to 2015 is an error in the wavelength calibration) exhibit no significant changes in their \Lya~profile shapes and total fluxes. For GJ 1214, \Lya~was not detected in 2011, and the observed flux (2.4\,$\times$\,10$^{-15}$ erg cm$^{-2}$ s$^{-1}$) is a 1-$\sigma$ upper limit determined by integrating the spectrum over the wavelength range 1214.4\,--\,1216.9 \AA. The 2015 observation (2.3 hr on-source with the 52\arcsec\,$\times$\,0.1\arcsec~slit) is more sensitive than the 2011 observation (2 hr on-source with the 52\arcsec\,$\times$\,0.05\arcsec~slit). The larger slit size and longer integration resulted in a 5.5-$\sigma$ detection of 2.1\,$\times$\,10$^{-15}$ erg cm$^{-2}$ s$^{-1}$ for GJ 1214's observed flux, consistent with the 2011 upper limit.

The observed flux variations can be explained by activity cycles (year-to-decade timescales) and/or stellar rotation (day-to-week timescales). However, the time variations may be at least partially due to systematic effects. In comparing COS and STIS observations while creating the MUSCLES panchromatic spectra, \cite{Loyd} find systematic offsets between the two instruments in the tens of percent. The offsets can be explained by slit losses due to non-optimal centering of the target in the narrow STIS slit (52\arcsec \,$\times$\,0.1\arcsec~for G140M and 0.2\arcsec \,$\times$\,0.06\arcsec~for E140M) in contrast to the minimal losses introduced by the larger COS aperture (2.5\arcsec). Different observations may show different levels of accuracy in centering the target in the slit, despite acquisition peakup exposures, thereby producing artificial changes in the overall flux levels observed in the narrow-slit STIS G140M and E140M observations.

\subsection{Extreme-UV Spectrum} \label{sec:EUVSpectrum}

\begin{figure}[t]
\centering
\subfigure{\includegraphics[width=0.5\textwidth,angle=0]
   {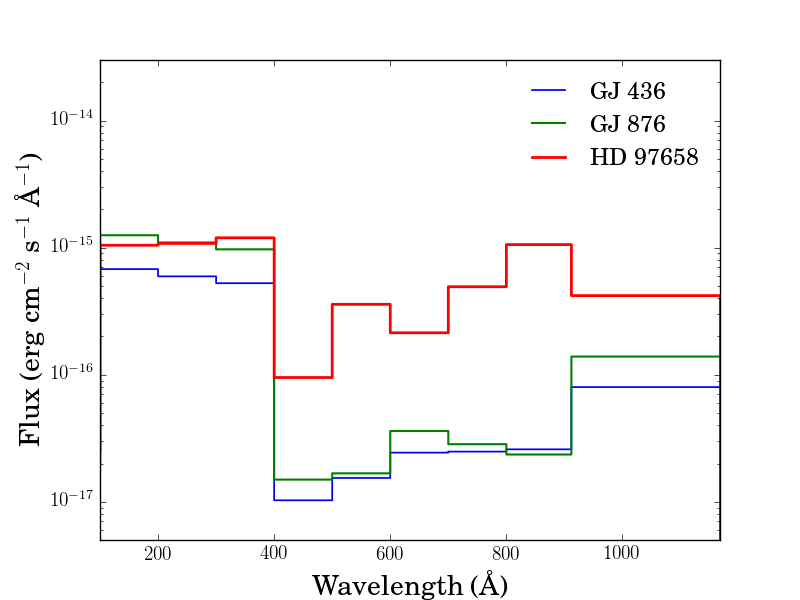}}
\caption{Extreme-UV spectra of GJ 436, GJ 876, and HD 97658 from 100\,--\,1170 \AA. These broadband spectra were constructed from the reconstructed stellar \Lya~fluxes using the empirical relations presented in \cite{Linsky2014}.}
\label{fig:EUV}
\end{figure}

We compute the extreme-UV spectrum for each target in nine ($\sim$100 \AA) bandpasses in the interval 100\,--\,1170 \AA~using empirically-derived relations between total \Lya~flux and extreme-UV flux presented by \cite{Linsky2014}. \cite{Linsky2014} used solar models ($\lambda$\,\textless \,2000 \AA) from \cite{Fontenla2013} to show that the ratio of extreme-UV flux to \Lya~flux varies slowly with the \Lya~flux. Using observations of F5 V\,--\,M5 V stars from $Far$ $Ultraviolet$ $Spectroscopic$ $Explorer$ ($FUSE$; 912\,--\,1170 \AA) and $Extreme$ $Ultraviolet$ $Explorer$ ($EUVE$; 100\,--\,400 \AA), and intrinsic (reconstructed) \Lya~fluxes from \cite{Wood2005} and \cite{Linsky2013}, they established $F$($\Delta \lambda$)/$F$(\Lya) ratios, where $\Delta \lambda$ represents one of the nine bandpasses in the interval 100\,--\,1170 \AA. The estimated accuracy of these relations (the rms dispersions about the fit lines) is $\sim$\,20\%. The extreme-UV fluxes ($\Delta$$\lambda$ = 100\,--\,1170 \AA) are presented in Table~\ref{table:LyAfluxes}, and the spectra are included in the High Level Science Products available on MAST. Example extreme-UV spectra for GJ 436, GJ 876, and HD 97658 are shown in Figure~\ref{fig:EUV}. In Section~\ref{sec:GJ436}, we compare our extreme-UV flux estimate for GJ 436 to the estimate from \cite{Ehrenreich2015} and find agreement within 30\%.

\section{Results} \label{sec:Results}

For each of the 11 target stars, the observed spectrum and its best-fit attenuated \Lya~model are shown in Figure~\ref{fig:all_MCMC}. See Tables~\ref{table:LyAfluxes} and \ref{table:Parameters} for the reconstructed \Lya~fluxes and best-fit parameters. In the following subsections, we discuss the fitting and results in greater detail. We use the local ISM models of \cite{Redfield2000,Redfield2008} to evaluate the plausibility of our ISM absorption best-fit parameters. For all 11 stars, we find \ion{H}{1}~velocity centroids that agree with the predictions of \cite{Redfield2008} within the MCMC uncertainties and STIS absolute velocity accuracies. The \cite{Redfield2000} model provides \ion{H}{1}~column density predictions only for the Local Interstellar Cloud (LIC), and our measured \ion{H}{1}~column densities do not always agree with the model predictions.

\subsection{GJ 832} \label{sec:GJ832}

GJ 832 is a M1.5 V star at 5.0 pc observed with the low-resolution G140M grating. GJ 832's sightline is predicted to have only one ISM component, the LIC with radial velocity -11.3\,$\pm$\,1.3 \kms~\citep{Redfield2008} and predicted log$_{10}$ \ion{H}{1}~column density 16.6 \citep{Redfield2000}.

The best-fit reconstructed \Lya~profile has a total flux $F$(\Lya) = (9.5\,$\pm$\,0.6)\,$\times$\,10$^{-13}$ erg cm$^{-2}$ s$^{-1}$, and the best-fit \ion{H}{1}~absorption has log$_{10}$ $N$(\ion{H}{1}) = 18.20\,$\pm$\,0.03 and velocity centroid $V_{\rm \HI}$ = -17.1$^{+0.8}_{-0.7}$ \kms~(Tables~\ref{table:LyAfluxes} and~\ref{table:Parameters}). The column density is much higher than the predicted value for the LIC. The LIC may be clumpier than \cite{Redfield2000}'s model could account for, but it is possible there is an unknown cloud component along this sightline. The best-fit velocity centroid is consistent with the prediction from \cite{Redfield2008} within G140M's absolute wavelength accuracy (2.5\,--\,6.2 \kms).

GJ 832's joint and marginalized probability distributions are shown in Figure~\ref{fig:GJ832_cornerplot} as the archetype for the high S/N (G140M or E140M) observations.  The marginalized distributions are all singly peaked, and the Doppler $b$ parameter's marginalized distribution is the only one exhibiting significant asymmetry.  Many joint distributions display degeneracies between parameters, and these are seen for all targets. See Section~\ref{sec:MCMC} for a description of the degeneracies.

\subsection{GJ 876} \label{sec:GJ876}

GJ 876 is a M5 V star at 4.7 pc observed with the low-resolution G140M grating. GJ 876's sightline is predicted to have only one ISM component, the LIC with radial velocity -3.6\,$\pm$\,1.4 \kms~\citep{Redfield2008} and predicted log$_{10}$ \ion{H}{1}~column density 17.7 \citep{Redfield2000}.

The best-fit reconstructed \Lya~profile has a total flux $F$(\Lya) = (3.9$^{+0.5}_{-0.4}$)\,$\times$\,10$^{-13}$ erg cm$^{-2}$ s$^{-1}$, and the best-fit \ion{H}{1}~absorption has log$_{10}$ $N$(\ion{H}{1}) = 18.03\,$\pm$\,0.04 and velocity centroid $V_{\rm \HI}$ = 2.8\,$\pm$\,0.8 \kms~(Tables~\ref{table:LyAfluxes} and~\ref{table:Parameters}). The column density is twice as large as the predicted value for the LIC, and the velocity centroid is consistent within G140M's absolute wavelength accuracy. As discussed in Section~\ref{sec:GJ832}, this large best-fit column density could indicate substructure within the LIC or an unknown cloud component along this sightline.

GJ 876's joint and marginalized distributions are singly peaked and similar in shape and symmetry to Figure~\ref{fig:GJ832_cornerplot}, including the asymmetry in the marginalized distribution for the Doppler $b$ value.

\subsection{GJ 436} \label{sec:GJ436}

GJ 436 is a M3.5 V star at 10.1 pc observed with the low-resolution G140M grating. According to the \cite{Redfield2008} model, GJ 436's sightline does not pass directly through any local cloud, but is close to or inside the edges of three clouds (the LIC, Leo, and NGP). \cite{Redfield2000} predicted a log$_{10}$ \ion{H}{1}~column density of 16.7 associated with the LIC.

The best-fit reconstructed \Lya~profile has a total flux $F$(\Lya) = (2.1\,$\pm$\,0.3)\,$\times$\,10$^{-13}$ erg cm$^{-2}$ s$^{-1}$, and the best-fit \ion{H}{1}~absorption has log$_{10}$ $N$(\ion{H}{1}) = 18.04\,$\pm$\,0.06 and velocity centroid $V_{\rm \HI}$ = -4.1$^{+1.3}_{-1.2}$ \kms~(Tables~\ref{table:LyAfluxes} and~\ref{table:Parameters}). The column density is consistent with other 10 pc distant stars reported in \cite{Wood2005}, indicating the presence of significant interstellar material.  

GJ 436's joint and marginalized distributions of the individual parameters are singly peaked and similar in shape and symmetry to Figure~\ref{fig:GJ832_cornerplot}, including the asymmetry in the marginalized distribution for the Doppler $b$ value. 

\cite{Bourrier2015} observed GJ 436 over two $HST$ visits and performed \Lya~reconstructions assuming an intrinsic \Lya~shape that allowed for self-reversal. Our $F$(\Lya) and log$_{10}$ $N$(\ion{H}{1}) values are in good agreement with theirs: $F$(\Lya) = 2.01\,$\times$\,10$^{-13}$ and 2.05\,$\times$\,10$^{-13}$ erg cm$^{-2}$ s$^{-1}$ and log$_{10}$ $N$(\ion{H}{1})\,=\,18.01\,$\pm$\,0.07 and 17.90$^{+0.15}_{-0.19}$.

We have also compared our extreme-UV flux computed using our reconstructed \Lya~flux (Section~\ref{sec:EUVSpectrum}) with the extreme-UV flux computed by \cite{Ehrenreich2015} using the X-ray flux from 5\,--\,100 \AA, via relations provided by \cite{Sanz-Forcada2011}. We find our extreme-UV fluxes are in satisfactory agreement. \cite{Ehrenreich2015} found $F$(EUV)\,=\,2.34\,$\times$\,10$^{-13}$ erg cm$^{-2}$ s$^{-1}$ over $\Delta \lambda$\,=\,124\,--\,912 \AA. Integrating our reconstructed extreme-UV over the same wavelength range, we find $F$(EUV)\footnote{The $F$(EUV)\,=\,2.1\,$\times$\,10$^{-13}$ erg cm$^{-2}$ s$^{-1}$ value reported in Table~\ref{table:LyAfluxes} is computed over $\Delta \lambda$\,=\,100\,--\,1170 \AA.}\,=\,1.7\,$\times$\,10$^{-13}$ erg cm$^{-2}$ s$^{-1}$, 27\% less than what \cite{Ehrenreich2015} found.

\subsection{GJ 581} \label{sec:GJ581}

GJ 581 is a M5 V star at 6.2 pc observed with the low-resolution G140M grating. GJ 581's sightline is predicted to have two ISM components: the G cloud with a radial velocity of -27.6\,$\pm$\,1.2  \kms~and the Gem cloud with radial velocity -23.8\,$\pm$\,1.0 \kms~\citep{Redfield2008}.

The best-fit reconstructed \Lya~profile has a total flux $F$(\Lya) = (1.1$^{+0.3}_{-0.2}$)\,$\times$\,10$^{-13}$ erg cm$^{-2}$ s$^{-1}$, and the best-fit \ion{H}{1}~absorption has log$_{10}$ $N$(\ion{H}{1}) = 18.01$^{+0.12}_{-0.17}$ and velocity centroid $V_{\rm \HI}$ = -24.1$^{+2.8}_{-1.7}$ \kms~(Tables~\ref{table:LyAfluxes} and~\ref{table:Parameters}). The velocity centroid is consistent with the predicted radial velocities within uncertainties for both clouds.

GJ 581's joint and marginalized distributions are singly peaked except for the Doppler $b$ value and exhibit a quality of shape and symmetry halfway between Figures~\ref{fig:GJ832_cornerplot} and \ref{fig:HD85512_cornerplot}.  Many of the parameters, while singly peaked, exhibit significant asymmetry in their marginalized distributions. This is reflected in the uncertainties listed in Tables~\ref{table:LyAfluxes} and~\ref{table:Parameters}.

\subsection{GJ 667C} \label{sec:GJ667C}
     
GJ 667C is a M1.5 V star at 6.8 pc observed with the low-resolution G140M grating. GJ 667C's sightline is predicted to have one ISM component, the G cloud with a radial velocity of -27.3\,$\pm$\,1.2 \kms~\citep{Redfield2008}.

The best-fit reconstructed \Lya~profile has a total flux $F$(\Lya) = (5.2\,$\pm$\,0.9)\,$\times$\,10$^{-13}$ erg cm$^{-2}$ s$^{-1}$, and the best-fit \ion{H}{1}~absorption has log$_{10}$ $N$(\ion{H}{1}) = 17.98$^{+0.12}_{-0.18}$ and velocity centroid $V_{\rm \HI}$ = -22.5$^{+2.3}_{-1.6}$ \kms~(Tables~\ref{table:LyAfluxes} and~\ref{table:Parameters}). The velocity centroid is consistent within G140M's absolute wavelength accuracy of the predicted radial velocity. 

GJ 667C's joint and marginalized distributions are singly peaked except for the Doppler $b$ value and exhibit at least a slight asymmetry for most of the parameters and significant asymmetry for the Doppler $b$ value.  These asymmetries are reflected in the uncertainties listed in Tables~\ref{table:LyAfluxes} and~\ref{table:Parameters}.

\subsection{GJ 176} \label{sec:GJ176}

GJ 176 is a M2.5 V star at 9.3 pc observed with the low-resolution G140M grating. GJ 176's sightline is predicted to have two ISM components: the LIC with radial velocity 23.7\,$\pm$\,0.9 \kms~and the Hyades cloud with radial velocity 12.9\,$\pm$\,1.1 \kms~\citep{Redfield2008}.

The best-fit reconstructed \Lya~profile has a total flux $F$(\Lya) = (3.9\,$\pm$\,0.2)\,$\times$\,10$^{-13}$ erg cm$^{-2}$ s$^{-1}$, and the best-fit \ion{H}{1}~absorption has log$_{10}$ $N$(\ion{H}{1}) = 17.46$^{+0.08}_{-0.10}$ and velocity centroid $V_{\rm \HI}$ = 29.0\,$\pm$\,0.4 \kms~(Tables~\ref{table:LyAfluxes} and~\ref{table:Parameters}). The velocity centroid is consistent with the predicted radial velocities of the LIC and Hyades cloud within G140M's absolute wavelength accuracy. \cite{Redfield2000} predicted that the LIC along GJ 176's sightline has a log$_{10}$ \ion{H}{1}~column density 18.2, a factor of 4 larger than our measurement. GJ 176's \ion{H}{1}~column density is among the lowest column densities measured for any star and is extremely low for a star at 9.4 pc. However, the best fit has an anomalously large Doppler $b$ value, 15.9\,$\pm$\,0.3 \kms. Forcing the $b$ value to a more probable 12 \kms~increases the \ion{H}{1}~column density by 30\%, but changes the reconstructed \Lya~flux by only a few percent. 

While we are skeptical of the fitted ISM parameter values, we have reproduced this solution with MPFIT and a gridsearch. GJ 176 is also the highest S/N observation of the MUSCLES sample, and its singly-peaked, symmetric joint and marginalized distributions indicate a good fit. If GJ 176 is among the sightlines with the lowest column densities, it would make an excellent target for future direct extreme-UV observations.

\begin{figure*}[t]
\centering
\subfigure{\includegraphics[width=\textwidth,angle=0]
   {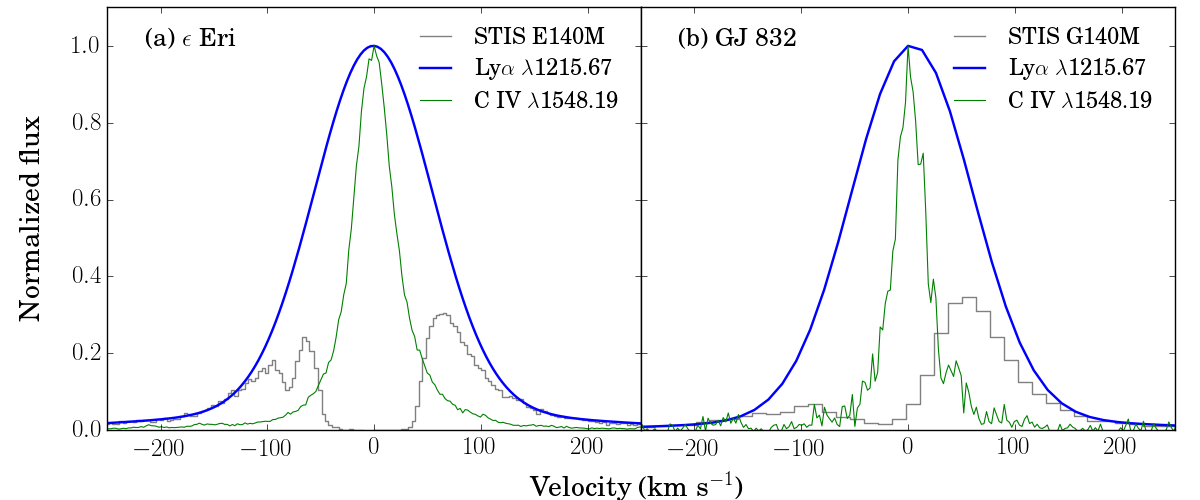}}
\caption{The thin gray histogram shows the observed \Lya~spectrum, the thick blue line shows the reconstructed \Lya~profile, and the thin green line shows the observed \CIV~($\lambda$1548) profile, plotted on a common velocity scale. We have defined 0 \kms~to be where each profile peaks.  (a) $\epsilon$~Eri.  (b) GJ 832. } 
\label{fig:LyaCIV_normprofiles}
\end{figure*}

\subsection{GJ 1214} \label{sec:GJ1214}
    
GJ 1214 is a M4.5 V star at 14.6 pc observed with the low-resolution G140M grating. GJ 1214's sightline is predicted to have two ISM components: the LIC with radial velocity -22.3\,$\pm$\,1.0 \kms~and the Mic cloud with radial velocity -26.6\,$\pm$\,1.0 \kms~\citep{Redfield2008}.

We estimate a total flux $F$(\Lya) = (1.3$^{+1.4}_{-0.5}$)\,$\times$\,10$^{-14}$ erg cm$^{-2}$ s$^{-1}$, and the best-fit \ion{H}{1}~absorption has log$_{10}$ $N$(\ion{H}{1}) = 18.06$^{+0.90}_{-0.42}$ and velocity centroid $V_{\rm \HI}$ = -26.4$^{+22.2}_{-82.0}$ \kms~(Tables~\ref{table:LyAfluxes} and~\ref{table:Parameters}). None of the parameters is well constrained. The predicted log$_{10}$ \ion{H}{1}~column density associated with the LIC along this sightline is 16.3, so the Mic cloud probably provides the dominant absorption. The best-fit \ion{H}{1}~velocity centroid agrees within 1\% of the predicted radial velocity of the Mic cloud, although its uncertainty is much larger.

In the MUSCLES pilot program \citep{France2013}, \Lya~emission was not detected from GJ 1214 with a 1-$\sigma$~upper limit 2.4 $\times$~10$^{-15}$ erg cm$^{-2}$ s$^{-1}$.  We now detect \Lya~at 5.5-$\sigma$~significance (2.1\,$\times$\,10$^{-15}$ erg cm$^{-2}$ s$^{-1}$), but the S/N is so low that we include only a single emission component in the model.  The fit is not well constrained, with the marginalized distributions not peaking for the column density and Doppler $b$ parameters over physically reasonable ranges (17.5\,$\leq$\,log$_{10}$ $N$(\ion{H}{1})\,$\leq$\,20; 1\,$\leq$\,$b$\,$\leq$\,20 \kms).  The other marginalized distributions, while singly peaked, exhibit large asymmetries as reflected in the reported uncertainties (Table~\ref{table:Parameters}).  

Because the fit quality is poor, we attempt to further constrain GJ 1214's intrinsic \Lya~emission using an empirical \Lya\,--\,\MgII~surface flux relation (see Section~\ref{sec:comparison}). Because these two transitions originate from gas with similar temperature and opacity, their fluxes are tightly correlated. Comparing the MCMC reconstructed flux to the flux expected from the empirical \Lya\,--\,\MgII~relation, we are likely underestimating GJ 1214's intrinsic \Lya~flux by a factor of 2.4. The best-fit line in Figure~\ref{fig:all_MCMC} (pink line) does appear to be an underestimate, and the orange line shows the scaled profile (the amplitude parameter has been increased by a factor of 2.4). The scaled profile's amplitude appears too large to match the observed profile, but the observed profile's flux is likely underestimated due to a possible over-subtraction of airglow. Thus, we present the scaled flux and individual parameters (only the amplitude has been modified (2.4$\times$) from the original MCMC fit) in Tables~\ref{table:LyAfluxes} and \ref{table:Parameters}.

\begin{figure*}[t]
\centering
\subfigure{\includegraphics[width=\textwidth,angle=0]
   {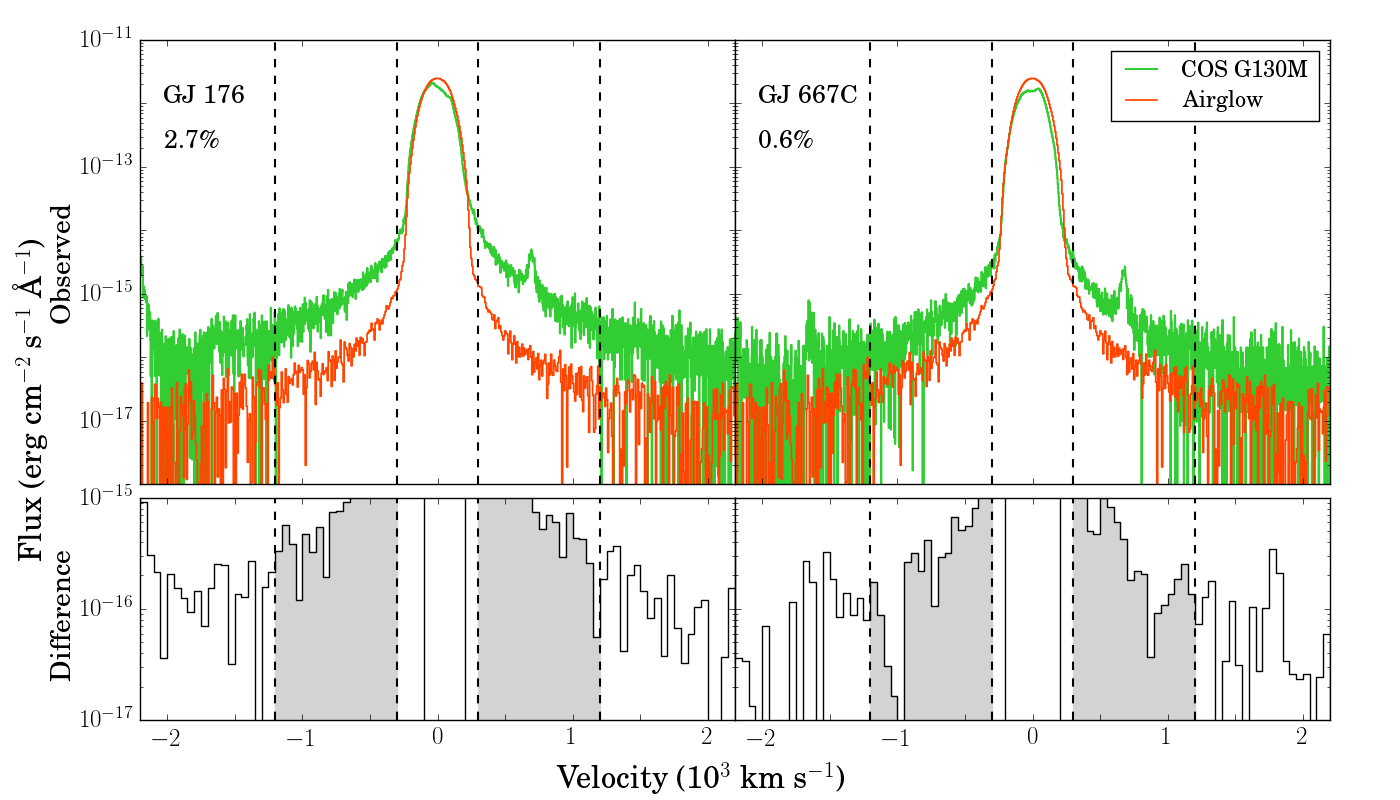}}
\caption{Broad \Lya~wings of two of our targets, GJ 176 and GJ 667C. The ``observed" flux panels show the \Lya~profiles observed with COS G130M (green) and the template COS G130M airglow spectrum (orange-red). The ``difference" flux panels show the COS G130M spectrum with the COS airglow template subtracted and the O\,\textsc{v} emission line at +650 \kms~removed. The dashed vertical lines and gray shading show the velocity ranges (-1200 to -300 \kms~and +300 to +1200 \kms) over which the ``difference" flux was integrated. The percentage of the total \Lya~flux contained in the gray-shaded regions is listed in each panel.}
\label{fig:HAWs_1}
\end{figure*}

\subsection{HD 85512} \label{sec:HD85512}

HD 85512 is a K6 V star at 11.2 pc observed with the medium-resolution E140M grating. HD 85512's sightline is predicted to have two ISM components: the G cloud with a radial velocity of -0.1\,$\pm$\,1.0 \kms~and the Cet cloud with radial velocity 14.3\,$\pm$\,0.7 \kms~\citep{Redfield2008}.

The best-fit reconstructed \Lya~profile has a total flux $F$(\Lya) = (1.2$^{+1.8}_{-0.5}$)\,$\times$\,10$^{-12}$ erg cm$^{-2}$ s$^{-1}$, and the best-fit \ion{H}{1}~absorption has log$_{10}$ $N$(\ion{H}{1}) = 18.43$^{+0.11}_{-0.14}$ and velocity centroid $V_{\rm \HI}$ = -3.3$^{+0.7}_{-0.8}$ \kms~(Tables~\ref{table:LyAfluxes} and~\ref{table:Parameters}). The velocity centroid is consistent with the G cloud, but not the Cet cloud. The Cet cloud may not significantly contribute to the absorption along this sightline. 

HD 85512's joint and marginalized distributions are shown in Figure~\ref{fig:HD85512_cornerplot} as the archetype of the E140M low S/N observations, although it is actually the worst case. The distributions are doubly peaked for half the parameters, and all but the \ion{H}{1}~velocity centroid exhibit significant asymmetry. The marginalized distribution of the broad component's velocity centroid is flat for values less than 40 \kms, so to not skew the best-fit value significantly, we limit the short wavelength end of the parameter range 50 \kms~further than for the narrow component's velocity centroid. We also limit the broad component's amplitude to 2 orders of magnitude fainter than the narrow component's range, because the MCMC cannot rule out a broad component consistent with zero flux. These requirements we had to place on the broad emission component's parameters indicate the broad component is only marginally detected.

\subsection{HD 97658} \label{sec:HD97658}

HD 97658 is a K1 V star at 21.1 pc observed with the medium-resolution E140M grating. HD 97658's sightline is predicted to have three ISM components: the LIC with predicted radial velocity of 2.7\,$\pm$\,1.4 \kms, the Leo cloud with predicted radial velocity 4.7\,$\pm$\,0.8 \kms, and the NGP with predicted radial velocity 9.5\,$\pm$\,0.8 \kms~\citep{Redfield2008}. \cite{Redfield2000} predicted a log$_{10}$ \ion{H}{1}~column density of 16.8 from the LIC along this sightline.

The best-fit reconstructed \Lya~profile has a total flux $F$(\Lya) = (9.1$^{+9.9}_{-4.5}$)\,$\times$\,10$^{-13}$ erg cm$^{-2}$ s$^{-1}$, and the best-fit \ion{H}{1}~absorption has log$_{10}$ $N$(\ion{H}{1}) = 18.45$^{+0.09}_{-0.14}$ and velocity centroid $V_{\rm \HI}$ = 7.1$^{+0.9}_{-1.0}$ \kms~(Tables~\ref{table:LyAfluxes} and~\ref{table:Parameters}). The velocity centroid is consistent within E140M's absolute wavelength accuracy (0.7\,--\,1.6 \kms) for the Leo and NGP clouds. The LIC is probably not the dominant source of absorption along this line of sight.

HD 97658's joint and marginalized distributions are singly peaked, except for the Doppler $b$ value, and exhibit at least slight asymmetry for most of the parameters.  These asymmetries are reflected in the uncertainties listed in Tables~\ref{table:LyAfluxes} and~\ref{table:Parameters}. The low S/N of this observation results in a large degeneracy between a low $b$ value (6 \kms), high column density (18.45) solution and a high $b$ value (12 \kms), low column density (18.2) solution. The low $b$ value, high column density fits the observed profile best, but the 12 \kms~is more realistic. Restricting the Doppler $b$ value to be \textgreater~9 \kms~results in $F$(\Lya) = (4.8$^{+3.7}_{-1.7}$)\,$\times$\,10$^{-13}$ erg cm$^{-2}$ s$^{-1}$, which is at the lower end of the original 1-$\sigma$ error bars.

\subsection{HD 40307} \label{sec:HD40307}

HD 40307 is a K2.5 V star at 13.0 pc observed with the medium-resolution E140M grating. HD 40307's sightline is predicted to have only one ISM component, the Blue cloud with a radial velocity of 7.6\,$\pm$\,1.1 \kms~\citep{Redfield2008}.

The best-fit reconstructed \Lya~profile has a total flux $F$(\Lya) = (2.0$^{+2.2}_{-0.9}$)\,$\times$\,10$^{-12}$ erg cm$^{-2}$ s$^{-1}$, and the best-fit \ion{H}{1}~absorption has log$_{10}$ $N$(\ion{H}{1}) = 18.60$^{+0.08}_{-0.09}$ and velocity centroid $V_{\rm \HI}$ = 9.4\,$\pm$\,1.0 \kms~(Tables~\ref{table:LyAfluxes} and~\ref{table:Parameters}). The velocity centroid is consistent within uncertainties with the predicted radial velocity.

HD 40307's joint and marginalized distributions are all singly peaked, although the Doppler $b$ value features a broad hump on its lower tail and the broad flux component's velocity centroid distribution becomes flat and non-zero at large velocities. This flat part of the distribution could significantly skew the median value away from the peak of the distribution, so we restricted the broad velocity centroid's parameter range extend 70 \kms~farther than the narrow component's range, because it seemed unlikely that the two flux components could be offset by such a large velocity. The necessity of this restriction indicates the broad emission component is only marginally detected. 

\subsection{$\epsilon$~Eri} \label{sec:epsEri}

$\epsilon$~Eri is a K2 V star at 3.2 pc observed with the medium-resolution E140M grating and is our highest S/N observation with that grating. $\epsilon$~Eri's sightline is measured to have only one ISM component, the LIC with radial velocity 20.9\,$\pm$\,1.8 \kms~\citep{Redfield2008} and column density 17.88\,$\pm$\,0.07 \citep{Dring1997}.

The best-fit reconstructed \Lya~profile has a total flux $F$(\Lya) = (6.1$^{+0.3}_{-0.2}$)\,$\times$\,10$^{-11}$ erg cm$^{-2}$ s$^{-1}$, and the best-fit \ion{H}{1}~absorption has log$_{10}$ $N$(\ion{H}{1}) = 17.93\,$\pm$\,0.02 and velocity centroid $V_{\rm \HI}$ = 16.1\,$\pm$\,0.4 \kms~(Tables~\ref{table:LyAfluxes} and~\ref{table:Parameters}). The velocity centroid is not quite consistent with the measured radial velocity, but the column density agrees with the measurement from \cite{Dring1997}. $\epsilon$~Eri's joint and marginalized distributions are all singly peaked and none display asymmetry.

$\epsilon$~Eri is the only star in our sample to show the presence of an astrosphere, as observed by \cite{Wood2005}.  To fit the attenuated profile, we mask the blue side of the \ion{H}{1}~absorption where the astrospheric absorption is present (Figure~\ref{fig:all_MCMC}) because the extra absorption skews the \ion{H}{1}~velocity centroid blueward and increases the Doppler $b$ value. The detected \ion{D}{1}~absorption eliminates the possibility of this blueward velocity centroid.

\section{Discussion} \label{sec:Discussion}

\begin{figure*}[t]
\centering
\subfigure{\includegraphics[width=\textwidth,angle=0]
   {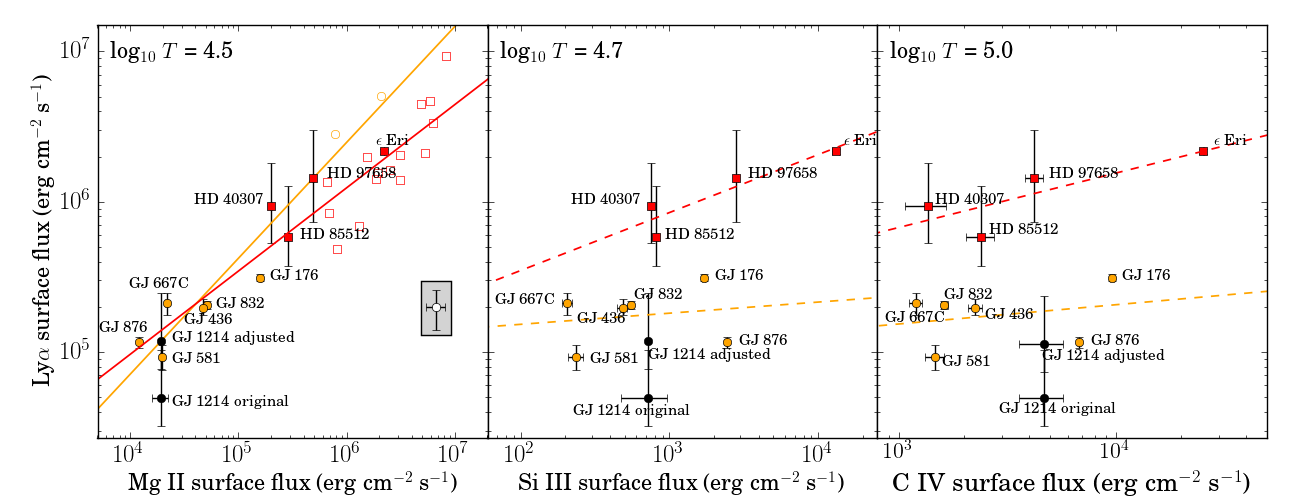}}
\caption{$Left:$ \Lya~surface flux versus \MgII~surface flux (corrected for ISM absorption). The M dwarfs are shown with orange circles and the K dwarfs in red squares (filled for the MUSCLES sample, open for the \cite{Wood2005} sample). GJ 1214's original and adjusted fluxes are shown with black circles because GJ 1214 was excluded from the fits. The best fit lines to the MUSCLES and \cite{Wood2005} data are shown in orange for the M dwarfs and red for the K dwarfs. For visual clarity, we show the size of the \cite{Wood2005} error bars in the grey shaded box. $Middle:$ \Lya~surface flux versus \SiIII~surface flux. $Right:$ \Lya~surface flux versus \CIV~surface flux. For the middle and right panels, the lines of best fit are shown as dashed lines because they are not statistically significant correlations (see Table~\ref{table:fitparms}). The formation temperatures of the metal ions are shown in the top left corner of each panel \citep{Dere1997,Landi2013}.}
\label{fig:LyaMgIISiIIICIV}
\end{figure*}

\begin{figure*}[t]
\centering
\subfigure{\includegraphics[width=\textwidth,angle=0]
   {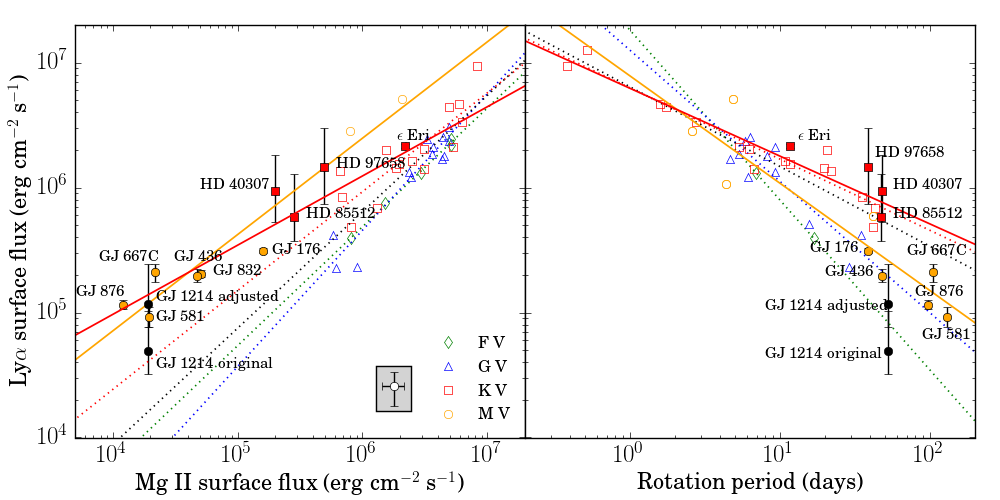}}
\caption{$Left:$ \Lya~surface flux versus \MgII~surface flux (corrected for ISM absorption). The M dwarfs are shown with orange circles and the K dwarfs in red squares (filled for the MUSCLES sample, open for the \cite{Wood2005} sample). GJ 1214's original and adjusted fluxes are shown with black circles because GJ 1214 was excluded from the least-squares fits. The best fit lines to the MUSCLES and \cite{Wood2005} data are shown in orange for the M dwarfs and red for the K dwarfs. The black dotted line is the least-squares fit to all of the \cite{Wood2005} data points, and the colored dotted lines are least-squares fits to the individual spectral types shown in the legend. For visual clarity, we show the size of the \cite{Wood2005} error bars in the grey shaded box. $Right:$ \Lya~surface flux versus stellar rotation period. Similar to the left panel, the solid points and lines correspond to data from the MUSCLES sample, and the open symbols and dotted lines correspond to data from \cite{Wood2005}. }
\label{fig:LyaMgII_WoodFig15}
\end{figure*}

\subsection{The \Lya~line profile} \label{sec:lineprofile}
The \Lya~emission lines observed by STIS have broad wings that are well-fit by a narrow and a broad Gaussian. Excluding our targets with poor S/N where the broad component is only marginally detected (HD 85512, HD 97658, HD 40307) or not at all (GJ 1214), the average and standard deviation of the narrow component FWHMs is 138\,$\pm$\,14 \kms, and 413\,$\pm$\,64 \kms~for the broad components. The broad flux component typically contributes only 10\% to the total intrinsic flux (Table~\ref{table:FluxRatios_HAWs}).

We compare the intrinsic \Lya~profiles with the \CIV~profiles ($\lambda$1550) and find that \Lya~is $\sim$\,3 times broader (Figure~\ref{fig:LyaCIV_normprofiles}). \CIV~has a higher formation temperature (log$_{10}$ $T$ = 5.0) than the \Lya~core (log$_{10}$ $T$ = 4.5) and forms only in the transition region while \Lya~forms in both the chromosphere and transition region. The abundance of neutral hydrogen is also much higher than triply-ionized carbon, and the measured ratio between the two members of the C IV doublet is $\sim$\,2. This confirms that \CIV~is optically thin, while photons must scatter out of the optically thick \Lya~line core to escape the stellar atmosphere.

The COS G130M \Lya~spectra, while overwhelmed by airglow in the inner $\pm$\,300 \kms~of the line core, are more sensitive to the broad wings of the \Lya~profile than are the STIS spectra. We detect broad \Lya~wings out to $\pm$\,1200 \kms. This flux probably originates from photons diffusing out of the \Lya~line core. To estimate how much of the total flux is in the broad wings (beyond $\pm$\,300 \kms), we subtract a template COS G130M airglow spectrum from the observed G130M spectrum, interpolate over the O\,\textsc{v} emission line at +650 \kms, and integrate the flux between $\pm$(300 -- 1200) \kms~(Figure~\ref{fig:HAWs_1}). We find percentages of the total \Lya~flux in the wings as high as 2.7\% (GJ 176) and as low as 0\% (GJ 1214) with an average of 1.1\% (Table~\ref{table:FluxRatios_HAWs}). However, the percentage flux in the broad wings depends on the STIS observation's S/N. As the S/N increases, the MCMC finds larger FWHM values for both the narrow and broad flux components, because the decrease in the noise floor allows for better sampling of the profile at large velocities. The S/N dependence indicates that the percentages listed in Figure~\ref{fig:HAWs_1} are lower limits, thus we estimate that \textgreater \,1\% of the total \Lya~flux comes from the $\pm$(300 -- 1200) \kms~wings of the MUSCLES targets.

\subsection{Comparison with other chromospheric and transition region lines} \label{sec:comparison}

Correlations between reconstructed \Lya~fluxes and other stellar emission lines will provide future studies with an alternative to estimate the intrinsic \Lya~flux and can aid the development of low-mass stellar atmospheric models. We compare \Lya~to metal ions of chromospheric or transition region origin whose observed fluxes and formation temperatures are listed in Table~\ref{table:LyAfluxes}. Many of the observed lines are multiplets, and for all but \CII~(whose $\lambda$1334 line is attenuated by the ISM), we derive the total flux by integrating over all multiplet members. Both \MgII~($\lambda \lambda$2796,2803) lines lose $\sim$30\% of their flux to interstellar attenuation \citep{France2013}, assuming a typical log$_{10}$ $N$(\MgII)\,$\sim$\,13 for stars within 20 pc \citep{Redfield2002}, so we adjust the \MgII~flux values by +30\% (Table~\ref{table:LyAfluxes}). The COS and STIS G230L spectra of \MgII~do not provide sufficient resolution for profile reconstruction. We do not adjust the $F$(\MgII) correction factor based on the $N$(\ion{H}{1}) value measured for each sightline, because the Mg$^{+}$/H is not fixed for the local ISM. It is sensitive to abundance variations as well as local temperature and ionization conditions. The list of compared ions is: \CII~($\lambda$1335), \MgII~($\lambda \lambda$2796,2803), \SiIII~($\lambda \lambda$1206,1206), \SiIV~($\lambda \lambda$1394,1403), \HeII~($\lambda$1640), and \CIV~($\lambda \lambda$1548,1551).  

We expect that the surface fluxes of species with similar formation temperatures will correlate well \citep{Linsky2013,France2013}. Surface fluxes are calculated using the distances and stellar radii listed in Table~\ref{table:LyAfluxes}. Figure~\ref{fig:LyaMgIISiIIICIV} compares the surface fluxes of \Lya~and three ions: \MgII~(line center log$_{10}$ $T$ = 4.5), \SiIII~(log$_{10}$ $T$ = 4.7), and \CIV~(log$_{10}$ $T$ = 5.0). In the \MgII~panel, we include the M and K dwarf samples from \cite{Wood2005} in the fit. We fit a line to the data for both spectral types (best-fit equations printed Table~\ref{table:fitparms}), but exclude GJ 1214 from each fit, because its reconstructed \Lya~flux is highly uncertain. Because we are mixing data sets that have different error analyses, we use uniform weighting for all the data points during the fit. As expected, we find positive correlations with lines whose formation temperature is closer to \Lya's line center log$_{10}$ $T$ = 4.5. Some scatter is expected due to time variability (the ions' spectra were taken close in time to \Lya's, but are not simultaneous) and slight differences or ranges in formation temperature. 

Here we evaluate the statistical significance of trends revealed in Figure~\ref{fig:LyaMgIISiIIICIV}. First, the \Lya\,--\,\MgII~correlations are the only ones with statistical significance ($p$-value\,\textless\,0.05) (Table~\ref{table:fitparms}). The slopes of the K- and M-dwarf \Lya\,--\,\MgII~best-fit lines are statistically consistent with each other, as are the K- and M-dwarf slopes for the \SiIII~and \CIV~relations shown in Figure~\ref{fig:LyaMgIISiIIICIV}. Therefore, we cannot conclude that our sample shows a steeper \Lya\,--\,\MgII~correlation for the M dwarfs than the K dwarfs. Figure~\ref{fig:LyaMgIISiIIICIV} shows that the M-dwarf best-fit lines for \SiIII~and \CIV~have shallower slopes compared to the M-dwarf \Lya\,--\,\MgII~correlation, but these fits are not statistically significant, probably due to our small sample size (4 K dwarfs and 6 M dwarfs; GJ 1214 was excluded from fitting). We do not show the fitting results for \CII, \SiIV, or \HeII, but the fit qualities (slope, $p$-value, $\sigma$) are similar to those for \SiIII~and \CIV. This might seem surprising for \CII, because of its similar formation temperature to \Lya, but \cite{Oranje1986} noted that \CII~behaves like a transition region line in M dwarfs. \cite{Fontenla2016} provide support of \CII~as a transition region line with their semi-empirical model of GJ 832's atmosphere, which shows that the transition region begins at temperatures $\sim$2000 K cooler than in the quiet Sun. 

We find no significant correlation between the chromospheric line \Lya~and the transition region emission lines \SiIII, \SiIV, \CII, \CIV, and \HeII, but a larger sample size is needed to place a strong constraint on a non-correlation. \cite{Oranje1986}, \cite{Schrijver1987}, and \cite{Rutten1989} showed that M dwarfs show a deficiency in chromospheric emission compared to FGK stars, but equivalent amounts of transition region and coronal emission compared to FGK stars. This deficiency appears independent of stellar magnetic activity, but may increase toward later-type stars with cooler effective temperature and higher surface gravity \citep{Rutten1989}. A shallow-sloped correlation (compared to the slope of a correlation between two chromospheric lines) between a chromospheric and transition region line has only been shown for optically-active M dwarfs (e.g., \citealt{Oranje1986}; albeit with sample sizes similar to the MUSCLES M-dwarf sample).

We use the \Lya\,--\,\MgII~best-fit line (Figure~\ref{fig:LyaMgIISiIIICIV}) to correct GJ 1214's \Lya~flux.  As discussed in Section~\ref{sec:GJ1214}, GJ 1214 is a weak \Lya~emitter and its low S/N spectrum may suffer from over-subtraction of airglow. By eye, it appears that our solution underestimates the \Lya~flux, and we are not confident in the robustness of its \Lya~profile reconstruction. Using the linear fit to the \Lya\,--\,\MgII~correlation (which excludes GJ 1214), we increase the total flux of GJ 1214 by a factor of 2.4. Figure~\ref{fig:LyaMgIISiIIICIV} shows the MCMC best-fit reconstructed flux for GJ 1214 labeled as ``original" and the scaled flux labeled as ``adjusted". 

In Figure~\ref{fig:LyaMgII_WoodFig15} we compare our observed \Lya\,--\,\MgII~and \Lya\,--\,rotation period correlations to those that \cite{Wood2005} found for stars of spectral classes F V\,--\,M V. The MUSCLES targets probe a new parameter space of lower-mass and less-active stars. The addition of the MUSCLES K dwarfs to \cite{Wood2005}'s K dwarf sample skews the relation to expect greater \Lya~fluxes for given \MgII~flux on the low-activity end of the parameter space. However, this difference in slope is not statistically significant, and the low-activity part of the parameter space has not been sampled with K-dwarf observations. \cite{Wood2005}'s sample of four M dwarfs had only two with both \Lya~and \MgII~flux measurements (AD Leo and AU Mic), and they could not create a M dwarf \Lya\,--\,\MgII~relation. The addition of these two stars to the MUSCLES sample nearly doubles the slope of the \Lya\,--\,\MgII~relation from what it would have been otherwise, but this difference is not statistically significant. 

The addition of the MUSCLES K dwarfs to \cite{Wood2005}'s K dwarfs does not significantly change the \Lya\,--\,rotation period relation. The entire \cite{Wood2005} M-dwarf sample (4 stars) has rotation period measurements, and the addition of the MUSCLES M dwarfs (6 stars excluding GJ 1214) allows the calculation of a \Lya\,--\,rotation period correlation. The difference in slope between K and M dwarfs of the \Lya\,--\,rotation period correlations is statistically significant at the the 2.4-$\sigma$ level. The correlations suggest that slowly-rotating M dwarfs will exhibit less \Lya~flux than slowly-rotating K dwarfs, but beyond the cross-over point of the relation, a faster-rotating M dwarf may exhibit more \Lya~flux than a similarly rotating K dwarf. However, this part of the parameter space has not been observationally explored for M dwarfs.

In Section~\ref{sec:EUVSpectrum}, we utilized empirical relations from \cite{Linsky2014} to estimate the broadband extreme-UV spectra from the reconstructed, intrinsic \Lya~fluxes for the MUSCLES sample. The \Lya~and extreme-UV fluxes should be positively correlated, because they both depend on the magnetic heating rate. Models of solar quiet and active regions show similarly-shaped temperature-pressure profiles \citep{Fontenla2013}, so all flux diagnostics should scale together, although not necessarily at the same rate. \cite{Claire2012} showed that the fraction of \Lya~photons with respect to all photons at $\lambda$\,\textless \,1700 \AA~is $\sim$\,40\% throughout the Sun's history, and this supports \Lya~flux as a good predictor of the extreme-UV flux. Much of the 100\,--\,1170 \AA~wavelength range contains \ion{H}{1}~and \ion{He}{1}~continua and emission lines with similar formation temperatures to \Lya. But, \Lya~is probably not a good predictor of the higher formation temperature transition region and coronal emission lines that dominate the 100\,--\,400 \AA~region of the extreme-UV, as seen by the poor correlations with transition region ions such as \CIV. However, extreme-UV reconstructions based on coronal X-rays and reconstructions based on chromospheric \Lya~agree well, within $\sim$\,30\% (Section~\ref{sec:GJ436}).

For the Sun, \cite{DeWit2005} did not find that \Lya~is the best emission line to predict the extreme-UV spectrum. If using only one or two emission lines to predict the solar extreme-UV spectrum, they recommend using \ion{O}{1}~($\lambda$1304 \AA), \CIV~($\lambda$1550 \AA), or \SiII~($\lambda$1817 \AA), or \Lya~in conjunction with \SiII~($\lambda$1817 \AA) or \ion{O}{1}~($\lambda$1304 \AA). Models of the chromospheres, transition regions, and coronae of M and K dwarfs, using the line fluxes provided by the MUSCLES survey, will be required to identify more robust scaling relations for low-mass stars.

\section{Summary} \label{sec:Summary}

Using an MCMC technique, we have reconstructed the intrinsic \Lya~profiles from the observed, attenuated profiles for 11 low-mass exoplanet host stars, and used the intrinsic \Lya~fluxes to estimate the extreme-UV spectra. We assume an intrinsic \Lya~profile consisting of a dominant narrow Gaussian and a weak broad Gaussian, and a single interstellar absorption component characterized by a Voigt profile. Our results are summarized as follows:

\renewcommand{\labelitemi}{--}

\begin{itemize}
  \item The intrinsic \Lya~profiles are broad ($\sim$\,300 \kms) with wings that extend to about $\pm$1200 \kms~($\Delta \lambda$~$\approx$~5 \AA). The vast majority of the flux is in the profile's narrow component; the wings beyond $\pm$\,300 \kms~contain at least 1\% of the total flux. 
  \item The intrinsic \Lya~surface flux correlates positively with \MgII, another chromospheric emission line, for both K and M dwarfs. There may be positive correlations of the \Lya~surface flux with lines formed at higher temperatures in the stellar atmosphere (e.g., \SiIII, \CIV), but our K and M dwarf samples are too small to make a statistically significant statement at this time.
  \item We have found new \Lya\,--\,\MgII~surface flux and \Lya\,--\,rotation period relations for K and M spectral types that span active and inactive stars. The relations change between spectral type probably owing to different temperature-pressure profiles in the stellar atmospheres.
  \item We measure \ion{H}{1}~column densities for 10 new sightlines ranging from 4.7\,--\,21.1 pc distant. The lower spectral resolution of the G140M data does not influence \ion{H}{1}~column densities inferred from the higher-resolution E140M spectra. Log$_{10}$ \ion{H}{1}~column densities range from 18.0\,--\,18.6 with one sight line, GJ 176, having an extremely low log$_{10}$ $N$(\ion{H}{1}) = 17.5. GJ 176 would make an excellent target for future extreme-UV observations.
\end{itemize}

\acknowledgements
The data presented here were obtained as part of the HST Guest Observing programs \#12464 and \#13650 as well as the COS Science Team Guaranteed Time programs \#12034 and \#12035. This work was supported by NASA grants HST-GO-12464.01 and HST-GO- 13650.01 to the University of Colorado at Boulder. We thank Rebecca Nevin and Evan Tilton for helpful discussions that shaped our methodology, and we thank the anonymous referee for suggestions that improved the discussion of the emission line correlations. Sarah Rugheimer would like to acknowledge support from the Simons Foundation (339489, Rugheimer). 

This research made use of the Python packages \texttt{emcee} \citep{Foreman-Mackey2013}, \texttt{pyspeckit} \citep{Ginsburg2011}, \texttt{lyapy} (https://github.com/allisony/lyapy), \texttt{scicatalog} (https://github.com/parkus/scicatalog), and \texttt{triangle} \citep{Foreman-Mackey2014}.

\bibliography{LyA_reconstruction_arxiv.bbl}{}
\bibliographystyle{apj}

\clearpage

\begin{turnpage}

\begin{deluxetable}{lcccccccccc} 
\tablecolumns{11} 
\tablewidth{0pt} 
\tablecaption{Emission Line Fluxes of MUSCLES Targets  \label{table:LyAfluxes}} 
\tablehead{ \colhead{Target} & 
                   \colhead{$d$} & 
                   \colhead{$R$} & 
                   \colhead{$F$(\Lya)} $^{a}$ & 
                   \colhead{$F$(EUV) $^{b}$} & 
                   \colhead{$F$(\MgII) $^{c}$} & 
                   \colhead{$F$(\SiIII)} & 
                   \colhead{$F$(\CII) $^{d}$} & 
                   \colhead{$F$(\SiIV)} & 
                   \colhead{$F$(\HeII)} & 
                   \colhead{$F$(\CIV)} \\ 
                   \colhead{Star}  & 
                   \colhead{(pc)} & 
                   \colhead{(R$_{\odot}$)} & 
                   \colhead{(erg cm$^{-2}$ s$^{-1}$)} & 
                   \colhead{(erg s$^{-1}$)} & 
                   \colhead{(erg cm$^{-2}$ s$^{-1}$)} & 
                   \colhead{(erg cm$^{-2}$ s$^{-1}$)} & 
                   \colhead{(erg cm$^{-2}$ s$^{-1}$)} & 
                   \colhead{(erg cm$^{-2}$ s$^{-1}$)} & 
                   \colhead{(erg cm$^{-2}$ s$^{-1}$)} & 
                   \colhead{(erg cm$^{-2}$ s$^{-1}$)} 
                   }
\startdata
$\mathrm{\epsilon}$ Eri & 3.2 & 0.74 $^t$ & (6.1$^{+0.2}_{-0.2}$)$\times$10$^{-11}$ & 5.3$\times$10$^{-11}$ & (6.28$\pm$0.02)$\times$10$^{-11}$ & (3.7$\pm$0.2)$\times$10$^{-13}$ & (4.51$\pm$0.01)$\times$10$^{-13}$ & (3.12$\pm$0.01)$\times$10$^{-13}$ & (5.57$\pm$0.03)$\times$10$^{-13}$ & (7.19$\pm$0.03)$\times$10$^{-13}$ \\[3pt]
HD 85512 & 11.2 & 0.7 $^u$ & (1.2$^{+1.8}_{-0.5}$)$\times$10$^{-12}$ & 6.8$\times$10$^{-13}$ & (6.0$\pm$0.1)$\times$10$^{-13}$ & (1.7$\pm$0.1)$\times$10$^{-15}$ & (4.0$\pm$0.1)$\times$10$^{-15}$ & (1.6$\pm$0.1)$\times$10$^{-15}$ & (3.0$\pm$0.9)$\times$10$^{-15}$ & (5.1$\pm$0.8)$\times$10$^{-15}$ \\[3pt]
HD 40307 & 13.0 & 0.83 $^u$ & (2.0$^{+2.2}_{-0.9}$)$\times$10$^{-12}$ & 1.5$\times$10$^{-12}$ & (4.4$\pm$0.1)$\times$10$^{-13}$ & (1.7$\pm$0.1)$\times$10$^{-15}$ & (2.8$\pm$0.1)$\times$10$^{-15}$ & (1.3$\pm$0.1)$\times$10$^{-15}$ & (2.3$\pm$0.9)$\times$10$^{-15}$ & (3.0$\pm$0.6)$\times$10$^{-15}$ \\[3pt]
HD 97658 & 21.1 & 0.72 $^v$ & (9.1$^{+9.9}_{-4.5}$)$\times$10$^{-13}$ & 6.7$\times$10$^{-13}$ & (3.08$\pm$0.05)$\times$10$^{-13}$ & (1.8$\pm$0.1)$\times$10$^{-15}$ & (2.4$\pm$0.1)$\times$10$^{-15}$ & (1.3$\pm$0.1)$\times$10$^{-15}$ & (2.0$\pm$0.3)$\times$10$^{-15}$ & (2.6$\pm$0.3)$\times$10$^{-15}$ \\[3pt]
GJ 832 & 5.0 & 0.46 $^w$ & (9.5$^{+0.6}_{-0.6}$)$\times$10$^{-13}$ & 9.6$\times$10$^{-13}$ & (2.39$\pm$0.03)$\times$10$^{-13}$ & (2.5$\pm$0.1)$\times$10$^{-15}$ & (3.8$\pm$0.1)$\times$10$^{-15}$ & (3.2$\pm$0.1)$\times$10$^{-15}$ & (6.8$\pm$0.3)$\times$10$^{-15}$ & (7.5$\pm$0.3)$\times$10$^{-15}$ \\[3pt]
GJ 876 & 4.7 & 0.37 $^x$ & (3.9$^{+0.4}_{-0.4}$)$\times$10$^{-13}$ & 3.8$\times$10$^{-13}$ & (4.0$\pm$0.1)$\times$10$^{-14}$ & (8.1$\pm$0.1)$\times$10$^{-15}$ & (1.06$\pm$0.01)$\times$10$^{-14}$ & (8.0$\pm$0.1)$\times$10$^{-15}$ & (5.5$\pm$0.3)$\times$10$^{-15}$ & (2.27$\pm$0.04)$\times$10$^{-14}$ \\[3pt]
GJ 581 & 6.2 & 0.3 $^y$ & (1.1$^{+0.3}_{-0.2}$)$\times$10$^{-13}$ & 1.1$\times$10$^{-13}$ & (2.4$\pm$0.1)$\times$10$^{-14}$ & (2.9$\pm$0.4)$\times$10$^{-16}$ & (4.8$\pm$0.4)$\times$10$^{-16}$ & (4.1$\pm$0.6)$\times$10$^{-16}$ & (7.2$\pm$1.9)$\times$10$^{-16}$ & (1.8$\pm$0.2)$\times$10$^{-15}$ \\[3pt]
GJ 176 & 9.3 & 0.45 $^x$ & (3.9$^{+0.2}_{-0.2}$)$\times$10$^{-13}$ & 4.0$\times$10$^{-13}$ & (2.03$\pm$0.02)$\times$10$^{-13}$ & (2.1$\pm$0.1)$\times$10$^{-15}$ & (5.4$\pm$0.1)$\times$10$^{-15}$ & (2.2$\pm$0.1)$\times$10$^{-15}$ & (6.7$\pm$0.3)$\times$10$^{-15}$ & (1.22$\pm$0.03)$\times$10$^{-14}$ \\[3pt]
GJ 436 & 10.1 & 0.45 $^x$ & (2.1$^{+0.3}_{-0.3}$)$\times$10$^{-13}$ & 2.1$\times$10$^{-13}$ & (5.0$\pm$0.1)$\times$10$^{-14}$ & (5.2$\pm$0.4)$\times$10$^{-16}$ & (1.1$\pm$0.1)$\times$10$^{-15}$ & (6.3$\pm$0.6)$\times$10$^{-16}$ & (1.5$\pm$0.2)$\times$10$^{-15}$ & (2.4$\pm$0.2)$\times$10$^{-15}$ \\[3pt]
GJ 667C & 6.8 & 0.46 $^w$ & (5.2$^{+0.9}_{-0.9}$)$\times$10$^{-13}$ & 5.3$\times$10$^{-13}$ & (5.4$\pm$0.1)$\times$10$^{-14}$ & (5.1$\pm$0.4)$\times$10$^{-16}$ & (6.5$\pm$0.5)$\times$10$^{-16}$ & (7.6$\pm$0.6)$\times$10$^{-16}$ & (1.9$\pm$0.2)$\times$10$^{-15}$ & (3.0$\pm$0.2)$\times$10$^{-15}$ \\[3pt]
GJ 1214 $^d$ & 14.6 & 0.21 $^z$ & (1.3$^{+1.4}_{-0.5}$)$\times$10$^{-14}$ & 1.4$\times$10$^{-14}$ & (2.2$\pm$0.4)$\times$10$^{-15}$ & (8.0$\pm$2.8)$\times$10$^{-17}$ & (9.8$\pm$3.0)$\times$10$^{-17}$ & (4.3$\pm$3.4)$\times$10$^{-17}$ & (7.4$\pm$14.4)$\times$10$^{-17}$ & (5.2$\pm$1.2)$\times$10$^{-16}$ \\[3pt] \hline
log$_{10}$ $T$ $^{f}$ &  &  & 4.5 (core) & & 4.5 (core) & 4.7 & 4.5 & 4.9 & 4.9 & 5.0\\
\enddata
\tablenotetext{a}{Reconstructed value.}
\tablenotetext{b}{$F$(EUV) $\Delta \lambda$~= 100 -- 1170 \AA.}
\tablenotetext{c}{We applied a correction factor of 1.3 to $F$(\MgII) to correct for ISM attenuation (see Section~\ref{sec:comparison}).}
\tablenotetext{d}{$F$(\CII) is the only metal flux presented here that is not doublet integrated ($\lambda$1335 \AA~only).}
\tablenotetext{e}{ GJ 1214's \Lya~flux represents the scaled value determined from the \Lya\,--\,\MgII~surface flux relation.}
\tablenotetext{f}{All formation temperatures are from the CHIANTI database \citep{Dere1997,Landi2013}}
\tablenotetext{*}{(t) \citealt{Edvardsson1993}, (u) \citealt{Eker2015}, (v) \citealt{Bonfanti2015}, (w) \citealt{Kraus2011}, (x) \citealt{vonBraun2014}, (y) \citealt{Zhou2013}, (z) \citealt{Berta2011}} 
\end{deluxetable}

\end{turnpage}

\clearpage	

%%% Shout out to Peter Williams for these 2 lines!
\global\pdfpageattr\expandafter{\the\pdfpageattr/Rotate 90}
\global\pdfpageattr\expandafter{\the\pdfpageattr/Rotate 0}

\begin{deluxetable}{lcccccccccr} 
\tablecolumns{11} 
\tablewidth{0pt} 
\tablecaption{ Parameters from \Lya~Reconstructions of MUSCLES Targets$^{\dagger}$ \label{table:Parameters}} 
\tablehead{ 
            \colhead{Target} & 
            \colhead{$V_{\rm narrow}$} & 
            \colhead{$A_{\rm narrow} ^a$} & 
            \colhead{FWHM$_{\rm narrow}$} & 
            \colhead{$V_{\rm broad}$} & 
            \colhead{$A_{\rm broad} ^a$} & 
            \colhead{FWHM$_{\rm broad}$} & 
            \colhead{log$_{10}$ $N(\rm \HI)$} & 
            \colhead{$b_{\rm \HI}$} & 
            \colhead{$V_{\rm \HI}$} & 
            \colhead{$\chi^2_{\nu}$} \\[4pt]
            \colhead{}  & 
            \colhead{(\kms)} & 
            \colhead{($\times$10$^{-13}$)} & 
            \colhead{(\kms)} & 
            \colhead{(\kms)} & 
            \colhead{($\times$10$^{-13}$)}  & 
            \colhead{(\kms)} & 
            \colhead{(cm$^{-2}$)} & 
            \colhead{(\kms)} & 
            \colhead{(\kms)} & 
            \colhead{}  
   }
\startdata
$\mathrm{\epsilon}$ Eri & 13.6$^{+0.4}_{-0.4}$ & 937.4$^{+44.4}_{-41.8}$ & 129.3$^{+1.6}_{-1.6}$ & 11.7$^{+1.8}_{-1.8}$ & 51.5$^{+2.8}_{-2.6}$ & 400.1$^{+9.0}_{-8.9}$ & 17.93$^{+0.02}_{-0.02}$ & 12.1$^{+0.2}_{-0.2}$ & 16.1$^{+0.4}_{-0.4}$ & 1.9 \\[4pt]
HD 85512 & -9.2$^{+3.3}_{-4.0}$ & 22.7$^{+37.8}_{-9.5}$ & 112.7$^{+16.8}_{-23.1}$ & -29.2$^{+11.3}_{-31.5}$ & 1.3$^{+3.6}_{-1.1}$ & 197.6$^{+81.1}_{-32.3}$ & 18.43$^{+0.11}_{-0.14}$ & 6.3$^{+6.6}_{-1.4}$ & -3.3$^{+0.7}_{-0.8}$ & 0.8 \\[4pt]
HD 40307 & 17.3$^{+5.1}_{-4.3}$ & 42.9$^{+50.5}_{-21.5}$ & 102.2$^{+13.1}_{-11.4}$ & 41.3$^{+29.8}_{-12.3}$ & 2.1$^{+3.0}_{-1.4}$ & 167.0$^{+32.4}_{-21.4}$ & 18.60$^{+0.08}_{-0.09}$ & 12.7$^{+1.5}_{-2.6}$ & 9.4$^{+1.0}_{-1.0}$ & 0.7 \\[4pt]
HD 97658 & -1.1$^{+2.5}_{-3.1}$ & 20.4$^{+23.1}_{-10.5}$ & 99.0$^{+12.8}_{-10.3}$ & -37.6$^{+18.7}_{-50.8}$ & 0.3$^{+0.3}_{-0.2}$ & 252.6$^{+54.3}_{-42.5}$ & 18.45$^{+0.09}_{-0.14}$ & 6.0$^{+5.6}_{-1.3}$ & 7.1$^{+0.9}_{-1.0}$ & 0.8 \\[4pt]
GJ 832 & 11.4$^{+1.4}_{-1.3}$ & 15.1$^{+1.1}_{-1.0}$ & 133.4$^{+2.4}_{-2.5}$ & 26.2$^{+4.2}_{-4.0}$ & 0.49$^{+0.07}_{-0.06}$ & 385.9$^{+17.5}_{-16.5}$ & 18.20$^{+0.03}_{-0.03}$ & 8.7$^{+1.0}_{-1.8}$ & -17.1$^{+0.8}_{-0.7}$ & 1.3 \\[4pt]
GJ 876 & 7.8$^{+1.1}_{-1.1}$ & 5.5$^{+0.7}_{-0.6}$ & 146.8$^{+5.2}_{-5.1}$ & 6.1$^{+5.9}_{-5.7}$ & 0.23$^{+0.05}_{-0.04}$ & 407.7$^{+31.6}_{-26.8}$ & 18.03$^{+0.04}_{-0.04}$ & 9.0$^{+1.3}_{-2.7}$ & 2.8$^{+0.8}_{-0.8}$ & 1.4 \\[4pt]
GJ 581 & -2.6$^{+3.6}_{-2.8}$ & 1.9$^{+0.5}_{-0.4}$ & 133.0$^{+5.0}_{-5.8}$ & -38.8$^{+31.6}_{-53.5}$ & 0.03$^{+0.02}_{-0.01}$ & 509.4$^{+208.1}_{-124.6}$ & 18.01$^{+0.12}_{-0.17}$ & 12.0$^{+1.5}_{-5.5}$ & -24.1$^{+2.8}_{-1.7}$ & 0.7 \\[4pt]
GJ 176 & 29.8$^{+0.6}_{-0.6}$ & 4.6$^{+0.3}_{-0.3}$ & 169.1$^{+3.2}_{-3.2}$ & 31.4$^{+4.2}_{-4.1}$ & 0.26$^{+0.03}_{-0.03}$ & 498.2$^{+24.1}_{-22.2}$ & 17.46$^{+0.08}_{-0.10}$ & 15.9$^{+0.3}_{-0.3}$ & 29.0$^{+0.4}_{-0.4}$ & 1.2 \\[4pt]
GJ 436 & 2.6$^{+1.5}_{-1.4}$ & 3.5$^{+0.6}_{-0.5}$ & 126.7$^{+5.2}_{-5.3}$ & 0.7$^{+8.2}_{-7.6}$ & 0.12$^{+0.16}_{-0.04}$ & 309.0$^{+46.1}_{-36.6}$ & 18.04$^{+0.06}_{-0.06}$ & 8.6$^{+1.4}_{-2.4}$ & -4.1$^{+1.3}_{-1.2}$ & 0.9 \\[4pt]
GJ 667C & 12.7$^{+4.5}_{-3.1}$ & 8.7$^{+1.7}_{-1.6}$ & 126.7$^{+3.7}_{-3.9}$ & 22.1$^{+7.0}_{-6.9}$ & 0.28$^{+0.06}_{-0.05}$ & 383.0$^{+30.5}_{-26.9}$ & 17.98$^{+0.12}_{-0.18}$ & 10.6$^{+2.0}_{-5.0}$ & -22.5$^{+2.3}_{-1.6}$ & 1.1 \\[4pt]
GJ 1214 $^{b}$ & 32.4$^{+31.7}_{-21.2}$ & 0.18$^{+0.20}_{-0.06}$ & 166.0$^{+27.1}_{-27.3}$ & -- & -- & -- & 18.06$^{+0.90}_{-0.42}$ & 14.9$^{+3.8}_{-6.2}$ & -26.4$^{+22.2}_{-82.0}$ & -- \\[2pt]
\enddata
\tablenotetext{$\dagger$}{D/H fixed at 1.5$\times$10$^{-5}$. }
\tablenotetext{a}{erg cm$^{-2}$ s$^{-1}$ \AA$^{-1}$. }
\tablenotetext{b}{GJ 1214's amplitude represents the scaled value determined from the \Lya\,--\,\MgII~relation.}
\end{deluxetable}

\begin{deluxetable}{lccc}
\tablecolumns{3}
\tablewidth{0pt}
\tablecaption{\Lya~broad component strengths compared to total intrinsic \Lya~flux \label{table:FluxRatios_HAWs}} 
\tablehead{\colhead{Target} & 
                  \colhead{$F_{\rm broad}$/$F_{\rm total}$ $^a$} & 
                  \colhead{$F_{\rm broad wings}$/$F_{\rm total}$ $^b$} \\
                  \colhead{} &
                  \colhead{(\%)} &
                  \colhead{(\%)}
                  }
\startdata

$\mathrm{\epsilon}$ Eri & 15 & -- $^c$ \\[4pt]
HD 85512 & 9 & 0.9 \\[4pt]
HD 40307 & 7 & 0.7 \\[4pt]
HD 97658 & 4 & 1.1 \\[4pt]
GJ 832 & 9 & 1.3 \\[4pt]
GJ 876 & 10 & 2.2 \\[4pt]
GJ 581 & 7 & 0.9 \\[4pt]
GJ 176 & 14 & 2.7 \\[4pt]
GJ 436 & 8 & 1.0 \\[4pt]
GJ 667C & 9 & 0.6 \\[4pt]
GJ 1214 & 0 $^d$ & 0 \\[2pt]
\enddata
\tablenotetext{a}{Percentage of the total reconstructed \Lya~flux ($F_{\rm total}$) contained in the broad Gaussian component ($F_{\rm broad}$).}
\tablenotetext{b}{Percentage of the total reconstructed \Lya~flux ($F_{\rm total}$) contained in the broad wings ($F_{\rm broad wings}$ is determined over $\Delta$$\lambda$\,=\,$\pm$(300\,--\,1200 \kms) from the COS G130M data).}
\tablenotetext{c}{$\mathrm{\epsilon}$ Eri does not have COS G130M data. }
\tablenotetext{d}{GJ 1214 was fit with only 1 Gaussian component.}
\end{deluxetable}

\begin{deluxetable}{cccccc}
\tablecolumns{6}
\tablewidth{0pt}
\tablecaption{ Parameters for fits in Figures~\ref{fig:LyaMgIISiIIICIV} and \ref{fig:LyaMgII_WoodFig15} \label{table:fitparms}} 
\tablehead{\colhead{Equation} & 
                  \colhead{Spectral Type} & 
                  \colhead{$\alpha$$^{a}$} & 
                  \colhead{$\beta$$^{a}$} &
                  \colhead{$p$-value$^{b}$} &
                  \colhead{$\sigma$$^{c}$}
                  }
\startdata

log$_{10}$ $F$(\Lya) = $\alpha$~log$_{10}$ $F$(\MgII) + $\beta$ & M V & 0.77$\pm$0.10 & 1.77$\pm$0.52  & 1.0\,$\times$\,10$^{-4}$ & 0.029 \\
 & K V & 0.55$\pm$0.11 & 2.77$\pm$0.68 & 4.3\,$\times$\,10$^{-5}$ & 0.031 \\
log$_{10}$ $F$(\Lya) = $\alpha$~log$_{10}$ $F$(\SiIII) + $\beta$ & M V & 0.07$\pm$0.31 & 5.04$\pm$0.87 & 0.75 & 0.033 \\
 & K V & 0.39 & 4.77 & 0.086 & 0.015 \\
 log$_{10}$ $F$(\Lya) = $\alpha$~log$_{10}$ $F$(\CIV) + $\beta$ & M V & 0.13$\pm$0.35 & 4.80$\pm$1.21 & 0.63  & 0.033\\
 & K V & 0.36 & 4.74 & 0.19 & 0.021\\
log$_{10}$ $F$(\Lya) = $\alpha$~log$_{10}$ $P$ + $\beta$ & M V & -0.86$\pm$0.16 & 6.89$\pm$0.24 & 3.3\,$\times$\,10$^{-4}$ & 0.037\\
 & K V & -0.54$\pm$0.05 & 6.79$\pm$0.05 & 3.2\,$\times$\,10$^{-10}$ & 0.019 \\
\enddata
\tablenotetext{a}{Uncertainties were determined by the fit's covariance matrix. For the K-dwarf \SiIII~and \CIV~relations, 4 data points are insufficient to calculate a covariance matrix.}
\tablenotetext{b}{The $p$-value is the probability of finding the best-fit line assuming the null hypothesis (no correlation) is true.}
\tablenotetext{c}{The standard deviation of the data points about the best-fit line, normalized by the fit.}
\end{deluxetable}

\end{document}